%% file: main.tex
\documentclass[runningheads]{llncs}

\usepackage{eccv}

\usepackage{eccvabbrv}

\usepackage{graphicx}
\usepackage{booktabs}
\usepackage{multirow}
\usepackage{gensymb}

\usepackage[accsupp]{axessibility}  

\usepackage{hyperref}

\begin{document}

\title{Sparse Light Field Sampling Improves Casual 3D and 4D Reconstruction}

\titlerunning{Sparse Light Field Sampling Improves Casual 3D and 4D Reconstruction}

\author{Shamus Li\inst{1} \and
Ruiming Cao\inst{2} \and
Laura Waller\inst{3} \and \\
Kristina Monakhova\inst{1} \and
Sara Fridovich-Keil\inst{4}}

\authorrunning{S.~Li et al.}

\institute{Cornell University, USA \and
Adobe, USA \and
University of California, Berkeley, USA \and
Georgia Institute of Technology, USA\\
\email{shamus@cs.cornell.edu}}

\maketitle

\begin{figure}[!ht]
\centering
\includegraphics[width=\linewidth]{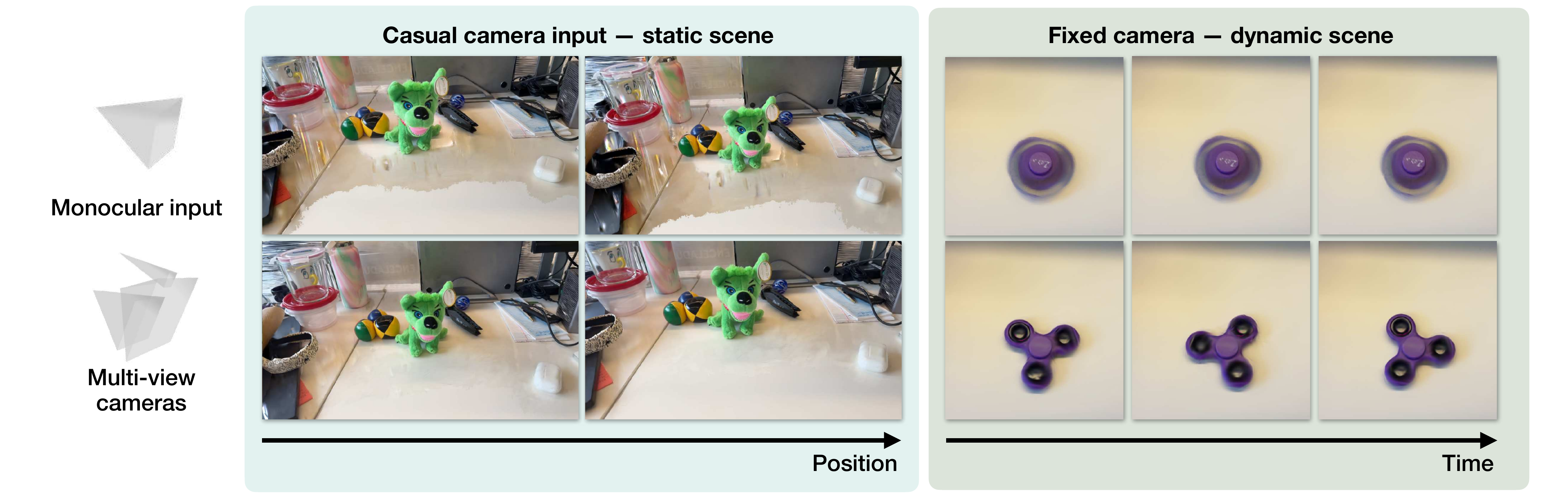}
\caption{\textbf{Sparse light field sampling.} In a casual-capture reconstruction of a static scene (\emph{left}), each exposure of a multi-view device records several viewpoints of the same scene state, recovering better geometry. Similarly, with a stationary camera observing a dynamic scene (\emph{right}), a monocular sensor records a single viewpoint per timestep and cannot distinguish motion from geometry, while synchronized cameras observe each scene state from multiple angles to break this ambiguity.}
\label{fig:teaser}
\end{figure}

\input{sec/0_abstract}

\input{sec/1_intro}
\input{sec/2_related_work}
\input{sec/3_benchmark}
\input{sec/4_experiments}

\input{sec/5_discussion}

\subsubsection*{Acknowledgements}
We thank Vi Tran for contributions on the multiplexed reconstruction pipeline, and Kevin C. Zhou for valuable discussions and assistance with the multiplexed light field hardware.

%
%
\bibliographystyle{splncs04}
\bibliography{main}

\end{document}


\title{Sparse Light Field Sampling Improves Casual 3D and 4D Reconstruction}
\subtitle{Supplementary Material}

\titlerunning{Sparse Light Field Sampling Improves Casual 3D and 4D Reconstruction}

\author{Shamus Li\inst{1} \and
Ruiming Cao\inst{2} \and
Laura Waller\inst{3} \and \\
Kristina Monakhova\inst{1} \and
Sara Fridovich-Keil\inst{4}}

\authorrunning{S.~Li et al.}

\institute{Cornell University, USA \and
Adobe, USA \and
University of California, Berkeley, USA \and
Georgia Institute of Technology, USA\\
\email{shamus@cs.cornell.edu}}

\maketitle
\appendix

\section{Background on Gaussian Splatting}

\subsection{3D Gaussian Splatting}
3D Gaussian Splatting (3DGS)~\cite{kerbl20233d} represents a scene as anisotropic 3D Gaussians. Gaussian $i$ has center $\boldsymbol{\mu}_i \in \mathbb{R}^3$, opacity $\eta_i$, view-dependent color $\mathbf{c}_i$, and covariance
\begin{equation}
    \boldsymbol{\Sigma}_i = \mathbf{R}_i\mathbf{S}_i\mathbf{S}_i^\top\mathbf{R}_i^\top,
\end{equation}
where $\mathbf{R}_i$ and $\mathbf{S}_i$ encode its rotation and scale. For a given camera, each Gaussian is projected to a 2D mean $\boldsymbol{\mu}^{\mathrm{2D}}_i$ and covariance $\boldsymbol{\Sigma}^{\mathrm{2D}}_i$. Its opacity contribution at pixel $\mathbf{x}$ is
\begin{equation}
    \alpha_i(\mathbf{x}) = \eta_i \exp\left[-\frac{1}{2}
    \left(\mathbf{x}-\boldsymbol{\mu}^{\mathrm{2D}}_i\right)^\top
    \left(\boldsymbol{\Sigma}^{\mathrm{2D}}_i\right)^{-1}
    \left(\mathbf{x}-\boldsymbol{\mu}^{\mathrm{2D}}_i\right)\right].
\end{equation}
After sorting the contributing Gaussians from front to back, 3DGS renders the pixel color using standard alpha compositing~\cite{max1995optical, mildenhall2020nerf}:
\begin{equation}
    \hat{\mathbf{C}}(\mathbf{x}) = \sum_{i=1}^N T_i(\mathbf{x})\alpha_i(\mathbf{x})\mathbf{c}_i,
    \qquad
    T_i(\mathbf{x}) = \prod_{j=1}^{i-1}\left(1-\alpha_j(\mathbf{x})\right).
\end{equation}
Thus, occlusion within each rendered sub-view is handled by the accumulated transmittance $T_i$.

\subsection{FSGS}
FSGS~\cite{zhu2024fsgs} adapts 3DGS to few-shot inputs primarily through a monocular depth prior. Rendered depth maps are supervised against the predictions of a pretrained monocular depth estimator using a scale-invariant depth correlation loss, which constrains scene geometry in regions that the sparse training views leave underdetermined. In addition, FSGS densifies the sparse SfM initialization through proximity-guided Gaussian unpooling, which places new Gaussians between existing neighbors to fill coverage gaps, and it synthesizes pseudo-views near the training cameras that are supervised with the same depth prior to regularize unseen viewpoints.

\subsection{SparseGS}
SparseGS~\cite{xiong2025sparsegs} likewise regularizes 3DGS with a pretrained monocular depth estimator, aligning rendered depth to predicted depth with a patch-based Pearson correlation loss that is robust to the unknown scale and shift of monocular predictions. It further incorporates a score distillation sampling term from a pretrained diffusion model to guide the appearance of novel viewpoints away from the training views, and a floater pruning step that detects and removes Gaussians responsible for foreground artifacts common in few-shot reconstruction.

\subsection{4D Gaussian Splatting}
4D Gaussian Splatting (4DGS)~\cite{wu20244d} extends 3DGS to dynamic scenes by introducing a deformation field that maps a set of canonical Gaussians $\mathcal{G} = \{ \mu, r, s, c, \alpha \}$ to their deformed states at a specific timestamp $t$. Rather than duplicating Gaussians for every frame---which incurs linear memory growth---4DGS learns a deformation function $\mathcal{F}$ to predict temporal residuals. To capture motion efficiently, 4DGS utilizes a K-planes factorization inspired by~\cite{cao2023hexplane, fridovich-keil2023kplanes}. The 4D spatiotemporal space is decomposed into six multi-resolution 2D feature planes: $\mathcal{P} = \{ \mathcal{R}_{xy}, \mathcal{R}_{xz}, \mathcal{R}_{yz}, \mathcal{R}_{xt}, \mathcal{R}_{yt}, \mathcal{R}_{zt} \}$.
For a canonical Gaussian at position $\mu = (x,y,z)$ and time $t$, a spatiotemporal feature vector $f_h$ is extracted via bilinear interpolation and fusion:
\begin{equation}
    f_h = \Phi_{\text{fusion}}\left( \bigcup_{\ell \in \mathcal{P}} \text{interp}(\mathcal{R}_\ell(\mu, t)) \right),
\end{equation}
where $\Phi_{\text{fusion}}$ is a lightweight MLP that aggregates features from the projected coordinate planes.

The aggregated feature $f_h$ serves as input to a small multi-head Gaussian deformation decoder $\mathcal{D}=\{\phi_\mu,\phi_r,\phi_s\}$, which consists of separate heads to predict residuals for position, rotation, and scaling:
\begin{equation}
    \Delta\mu = \phi_\mu(f_h), \quad \Delta r = \phi_r(f_h), \quad \Delta s = \phi_s(f_h).
\end{equation}
The deformed Gaussians $\mathcal{G}'_t = \{ \mu + \Delta\mu, r + \Delta r, s + \Delta s, c, \alpha \}$ are then rendered using the standard 3DGS renderer.

\section{Multiplexed Light Field Camera Details}

\subsection{Optical Design}
Placing a lenslet array in the Fourier plane of an imaging system is a common design in microscopy for 3D depth imaging~\cite{levoy2006light, pegard2016compressive, scrofani2018fimic, guo2019fourier}, and our design draws in particular on \cite{georgiev2006light} and \cite{liu2020fourier}. The proposed system consists of four main components: (i) a main objective lens, (ii) a lens array, (iii) an aperture array, and (iv) the image sensor. To maximize the angular diversity of rays entering the system, we use a large-aperture main lens as a main objective. For mesoscale imaging applications, this lens can be mounted in reverse to achieve high magnification, and the object should be placed close to the lens in order to maximize the perspective shift captured by the system. We place an $M\times N$ lens array in the system, which forms an array of sub-images at the sensor. In a standard plenoptic camera, the distance between the lens array and sensor is chosen such that these sub-images do not overlap. In our multiplexed design, we reduce this distance to increase the size of each sub-image. To control the trade-off between light throughput, depth of field, and optical crosstalk, we add an aperture array aligned with the lens array. This array blocks high-angle stray light that might otherwise skip adjacent lenslets and limits the effective aperture size of each sub-view, extending the depth of field for individual perspective images.

\subsection{Forward Model}
The system is modeled as an array of independent cameras. Each lenslet $(i,j)$ projects radiance from a unique viewing direction onto the sensor. The multiplexed sensor measurement $y(\mathbf{x})$ is a linear superposition of the images formed by each lenslet. Let $I_{ij}(\mathbf{x})$ denote the image formed by lens $(i,j)$ at sensor location $\mathbf{x}$, and let $w_{ij}(\mathbf{x})$ indicate whether that lens' footprint covers pixel $\mathbf{x}$. The multiplexed sensor measurement is:
\begin{equation}
    y(\mathbf{x}) = \frac{1}{M_{\text{max}}}\sum_{i,j}w_{ij}(\mathbf{x})I_{ij}(\mathbf{x})
\end{equation}
where $M_{\text{max}} = \max_{\mathbf{x}'}\sum_{i,j}w_{ij}(\mathbf{x}')$ is the maximum number of overlapping lenses across the image.

\subsection{Camera Calibration}
Since the sub-views are mixed in a single capture, we cannot use standard structure-from-motion calibration directly on the multiplexed image, nor can we extract the sub-images as in the traditional light field case. To calibrate our multiplexed system, we require a one-time calibration procedure in which we physically mask all lenslets except one, sequentially for all lenslets in the array. This yields a set of isolated sub-view images of a textured calibration object. These calibration images can then be processed using VGGT~\cite{wang2025vggt} with an additional bundle adjustment step to estimate intrinsic and extrinsic parameters. Once calibrated, the camera parameters are fixed and can be used for subsequent single-shot captures of new scenes. During dataset collection, the spatial masks $w_{ij}(\mathbf{x})$ for each lenslet can be determined by capturing white field images with uniform illumination.

\section{Additional Implementation Details}
\begin{figure}[H]
\centering
\includegraphics[width=0.7\linewidth]{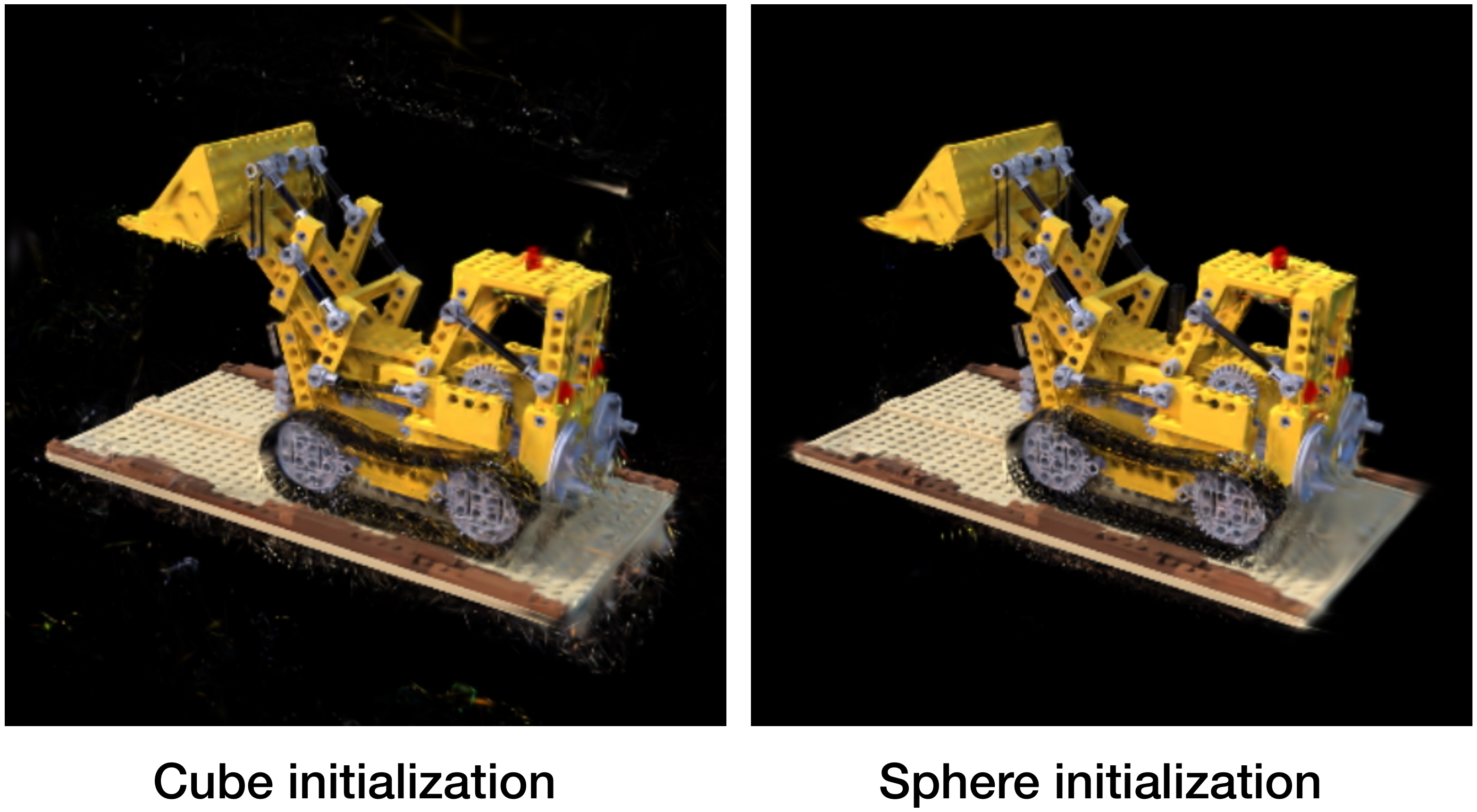}
\caption{Comparison between random cube sampling (left) and random sphere sampling (right) as 3D Gaussian initialization strategies. Sampling a sphere when initializing Gaussians reduces floater artifacts.}
\label{fig:ablation-initialization}
\end{figure}

Since the Blender scenes do not have an initial Structure-from-Motion (SfM)-derived point cloud, we follow 3DGS~\cite{kerbl20233d} and initialize 100K Gaussians by random sampling. Unlike 3DGS, however, we sample randomly within a sphere containing the target object, rather than a cube, as we empirically observe that this reduces floater artifacts, particularly in background regions. We note that~\cite{fridovich-keil2022plenoxels} made a similar choice for bounded scenes, by initializing voxel densities to zero outside a sphere containing the object. \cref{fig:ablation-initialization} compares uniform random sphere sampling against random cube sampling as initialization strategies. The spherical support more closely matches the bounded object-centric synthetic scenes and avoids placing Gaussians in the empty corners of the enclosing cube. We train for approximately 3,000 iterations, which is sufficient for convergence in our experiments.

For real-world experiments, where SfM point clouds are available from the calibration process described in the main paper, we initialize the Gaussians using these sparse points following 3DGS~\cite{kerbl20233d}.

\begin{figure}[H]
\centering
\includegraphics[width=\linewidth]{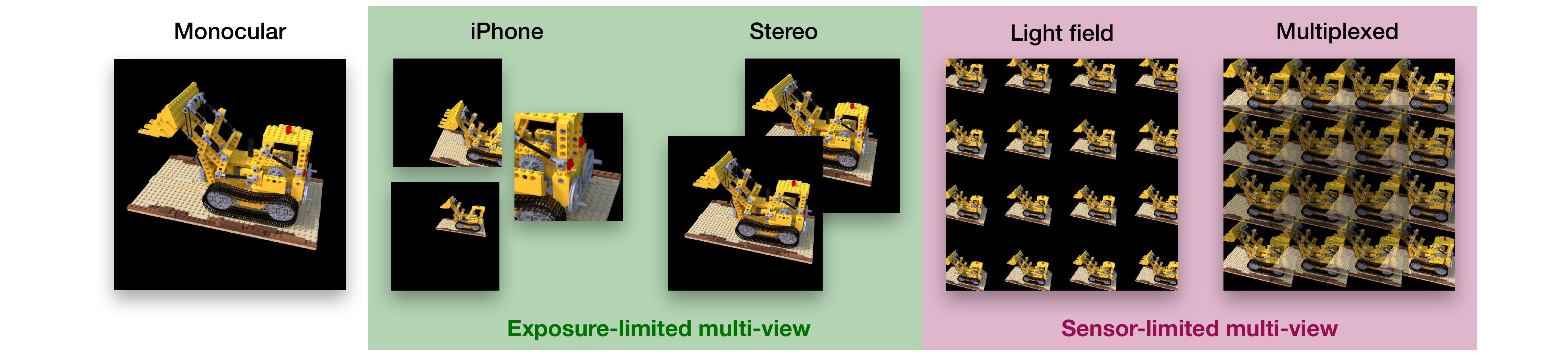}
\caption{Single-exposure measurements from each camera model. Sensor-limited cameras (light field, multiplexed) divide one fixed sensor between spatial and angular sampling, whereas exposure-limited cameras (iPhone, stereo) add full-resolution synchronized sensors per exposure.}
\label{fig:input-examples}
\end{figure}

\Cref{fig:input-examples} shows a single-exposure measurement of the \emph{lego} scene for each camera model in our benchmark.

\section{Additional Per-Scene Results}
We report per-scene metrics in \Cref{tab:static-synthetic-per-scene} (static synthetic scenes, sparse capture), \Cref{tab:static-real-per-scene} (static real scenes, single-exposure capture), \Cref{tab:casual-real-per-scene} (static real scenes, casual video capture), and \Cref{tab:dynamic-real-per-scene} (dynamic scenes, fixed camera).
\input{tables/static-synthetic-per-scene}
\input{tables/static-real-per-scene}
\input{tables/casual-real-per-scene}
\input{tables/dynamic-real-per-scene}

\FloatBarrier

%
%
{
\small
\bibliographystyle{splncs04}
\bibliography{main}
}

%% file: sec/0_abstract.tex
\begin{abstract}
Many consumer smartphones, stereo cameras, and light field cameras record multiple synchronized viewpoints in a single exposure event. However, novel view synthesis pipelines commonly use only a monocular stream and rely on camera motion or learned priors to obtain angular coverage. In this paper, we ask: why do we use only one viewpoint? We analyze \textit{sensor-limited multi-view}, where one sensor trades off spatial and angular resolution, and \textit{exposure-limited multi-view}, where multiple sensors on one commodity device observe each event simultaneously. We introduce a new dataset incorporating three types of commodity multi-view cameras, and evaluate sparse-view 3DGS and 4DGS baselines measuring reconstruction quality as a function of number of exposures and angle between extreme views. Our results demonstrate that using multiple cameras---even with a low baseline---significantly improves reconstruction quality in single-shot, few-shot, and casual video settings. In addition, under a fixed sensor budget, angular sampling improves reconstruction when exposures are scarce despite lower spatial resolution. The gains are most pronounced for single-shot and dynamic scenes, where a stationary monocular camera lacks the angular diversity to recover scene geometry and motion. Code, data, and additional results are available on our project page: \url{https://shamus.li/lightfield-gaussian-splatting}.

\keywords{Novel view synthesis \and Sparse-view 3D reconstruction \and Light field imaging \and Gaussian splatting \and Dynamic scene reconstruction}
\end{abstract}

%% file: sec/1_intro.tex
\section{Introduction} \label{sec:intro}
Most photographs and videos today are captured on devices with multiple cameras. For instance, the three rear cameras on an iPhone~\cite{iphone} simultaneously image a scene from viewpoints separated by roughly $5^\circ$; the stereo main cameras on an Apple Vision Pro~\cite{apple} provide a much wider baseline of approximately $15^\circ$; and even legacy light field cameras such as the Lytro Illum~\cite{ng2005light} record a $13 \times 13$ grid of sub-aperture views in a single shot. Such cameras capture angular diversity that a monocular sensor cannot provide.

However, the novel view synthesis community almost exclusively assumes monocular input. Methods built on NeRF~\cite{mildenhall2020nerf}, 3DGS~\cite{kerbl20233d}, and their dynamic counterparts~\cite{wu20244d, fridovich-keil2023kplanes, cao2023hexplane} assume a single image per camera pose. To accumulate angular diversity, existing approaches either build expensive multi-camera rigs~\cite{broxton2020immersive, li2022neural}, or move the camera around the scene, introducing an inherent tradeoff between angular and temporal resolution. Further, moving the camera only provides effective multi-view cues when the camera moves much faster than the scene, which is impractical for dynamic scenes~\cite{gao2022monocular}. When views are limited, recent works typically rely on learned priors~\cite{wu2023reconfusion, wu2025difix3d} or regularizers~\cite{niemeyer2022regnerf, yang2023freenerf} rather than better hardware. This is true even of datasets captured on multi-camera phones and tablets~\cite{baruch2021arkitscenes, gao2022monocular, yeshwanth2023scannet, ling2024dl3dv10k}: the devices have two or more cameras, but only one is used for training.

Why is this angular information discarded? Perhaps the most natural theory is that consumer camera baselines are too small to provide meaningful parallax cues, and thus no meaningful improvement over monocular capture. Our work tests and rejects this hypothesis. We frame the question through the lens of light field sampling~\cite{levoy1996light, gortler1996lumigraph}, and analyze two regimes: \textit{sensor-limited multi-view}, where one finite sensor trades off spatial and angular resolution, and \textit{exposure-limited multi-view}, where multiple synchronized sensors on one commodity device observe each exposure event simultaneously. A moving monocular camera would match the image count only by spending more capture time and, for dynamic scenes, observing a different scene state.

We conduct a comprehensive evaluation of Gaussian splatting and prior-based sparse-view methods across three types of consumer multi-view cameras---smartphone multi-camera arrays, stereo cameras, and light field cameras---and compare them against monocular baselines across single-shot, few-shot, and casual video captures of both static and dynamic scenes. We additionally propose a multiplexed light field camera, an extension of the light field camera's spatial-angular tradeoff in which higher-resolution sub-lens images intentionally overlap on the sensor.

We demonstrate that using all available cameras substantially improves few-shot static and dynamic reconstruction, as shown in \Cref{fig:teaser}. Under a fixed sensor budget, angular sampling improves reconstruction when exposures are scarce despite lower per-view spatial resolution, and the multiplexed design improves this tradeoff both in simulation and on our physical prototype. The gains are most pronounced for single-shot and dynamic scenes, where a stationary monocular camera lacks the angular diversity to recover scene geometry and motion. Moreover, these hardware-derived gains complement data-driven priors; the additional scene-specific angular information encoded in each exposure is not subsumed by the statistical information encoded by a learned prior. Concretely, our contributions are:
\begin{itemize}
    \item \textbf{Multi-camera datasets.} Static and dynamic real-world datasets in which each scene is captured by an iPhone~15~Pro (3 views per exposure), an Apple Vision~Pro (stereo), and a Lytro Illum light field camera (81 views per exposure).
    \item \textbf{Cross-camera benchmark.} A systematic evaluation of 3DGS, 4DGS, and sparse-view Gaussian splatting baselines across consumer multi-view cameras, measuring quality as a function of camera type, exposure count, and angular baseline.
    \item \textbf{Practical findings.} We show that (i) consumer cameras that sample the light field outperform monocular cameras in view-limited regimes, and (ii) the benefit is most pronounced for dynamic scenes, where angular diversity at each timestep is essential.
    \item \textbf{Hardware-software co-design.} We build a multiplexed light field camera prototype in which higher-resolution sub-lens images overlap on the sensor. On real captures, it outperforms both monocular and standard light field capture from a single exposure, illustrating how the angular-spatial tradeoff can be improved when camera hardware is designed in concert with the reconstruction algorithm.
\end{itemize}

%% file: sec/2_related_work.tex
\section{Related Work}
\label{sec:related-work}

\subsubsection{Sparse-view, casual capture, and dynamic novel view synthesis.}
Radiance field methods~\cite{mildenhall2020nerf, fridovich-keil2022plenoxels, kerbl20233d, barron2021mipnerf} assume a single monocular image per camera pose, requiring the user to move the camera to accumulate angular coverage. The problem becomes increasingly ill-posed when views are sparse, trajectories are short, or scenes are dynamic~\cite{gao2022monocular, zhao2022humannerf}. Approaches to novel view synthesis have largely focused on stronger priors: structural regularizers such as depth smoothness, total variation, and frequency constraints~\cite{niemeyer2022regnerf, yang2023freenerf, wang2023sparsenerf, xiong2025sparsegs, zhu2024fsgs}; data-driven and generative priors~\cite{yu2021pixelnerf, wu2023reconfusion, gao2024cat3d, wynn2023diffusionerf, liu2023zero1to3, wu2025difix3d}; and specialized dynamic representations~\cite{wu20244d, fridovich-keil2023kplanes, cao2023hexplane, pumarola2021dnerf, yang2024deformable, li2024spacetime, wang2025shape}. Feed-forward models~\cite{szymanowicz2024splatter, charatan2024pixelsplat, chen2021mvsnerf} can accept one or a few views, but assume these views come from camera motion rather than synchronized multi-camera capture. While priors are effective at constraining geometry toward plausible shapes, they remain fundamentally limited by insufficient angular measurements of the scene at each timestep. We evaluate both priors and additional sensing in our experiments, and report how they interact.

\subsubsection{Benchmarks and datasets.}
Standard novel view synthesis (NVS) benchmarks~\cite{mildenhall2020nerf, barron2022mipnerf, jensen2014large, knapitsch2017tanks, reizenstein2021common, gao2022monocular} all assume monocular input. Although several widely used datasets are collected on multi-camera phones and tablets~\cite{baruch2021arkitscenes, yeshwanth2023scannet, ling2024dl3dv10k}, they retain only a single camera stream. No existing NVS benchmark evaluates reconstruction quality as a function of angular sampling density, and to our knowledge, none compares multiple types of consumer cameras on the same scenes.

\subsubsection{Light field sampling.}
Standard monocular cameras capture a limited angular range with each exposure. Collecting additional angular measurements requires either moving the camera~\cite{gao2022monocular, tian2023mononerf}, exploiting mirrors placed around a small-scale scene~\cite{zhang2025seeing}, or purchasing multiple cameras and arranging them around a scene~\cite{li2022neural, isik2023humanrf, mildenhall2019local, wilburn2005high, broxton2020immersive}. Instead, other camera models sample multiple angles of the light field~\cite{levoy1996light, gortler1996lumigraph} with each exposure. Light field cameras~\cite{adelson1992single, ng2005light, georgiev2006light, lumsdaine2008full, perwass2012single} insert a microlens array between the sensor and main lens, increasing angular resolution at the cost of spatial resolution; compressive designs~\cite{marwah2013compressive, liang2008programmable} shift this tradeoff by multiplexing. Learning-based methods can synthesize novel views directly from such light field input~\cite{kalantari2016learningbased, suhail2022light}.

Plenoptic sampling theory~\cite{chai2000plenoptic, zhang2003spectral, davis2012unstructured, mildenhall2019local} establishes the angular sampling density needed for alias-free reconstruction, but the required rates typically demand dedicated hardware or dense capture trajectories. However, many consumer devices already perform sparse light field sampling without meeting these theoretical requirements: stereo cameras simultaneously capture two high-resolution views separated by a small baseline~\cite{apple, lee2024generalizable}, dual-pixel sensors provide two sub-aperture images~\cite{punnappurath2020modeling, garg2019learning}, and modern smartphone cameras include several sub-cameras with different focal lengths~\cite{iphone}. Whether this sparse angular sampling---far below the density needed for alias-free reconstruction---offers substantial benefits has not been systematically evaluated. Our analysis addresses this gap.

\begin{figure}[!b]
\centering
\includegraphics[width=\linewidth]{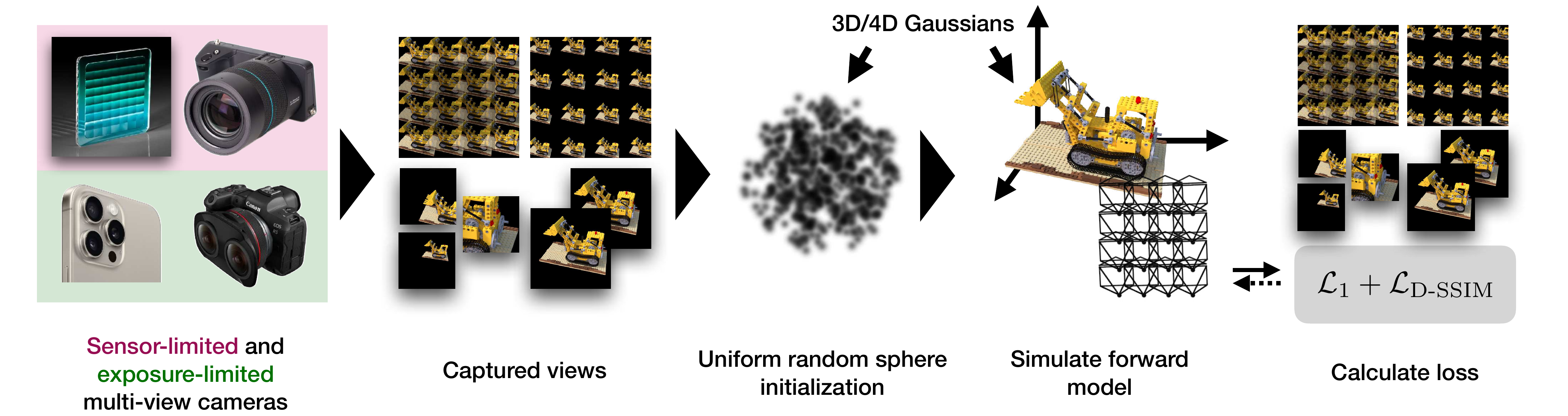}
\caption{\textbf{Method overview.} We evaluate 3DGS~\cite{kerbl20233d} and 4DGS~\cite{wu20244d} on sensor-limited and exposure-limited multi-view cameras. Each camera is modeled as an array of sub-views with known relative poses, and all sub-views of an exposure supervise the same set of Gaussians through the camera's forward model.}
\label{fig:method}
\end{figure}

%% file: sec/3_benchmark.tex
\section{Light Field Sampling Cameras}
\label{sec:benchmark}

Classic plenoptic sampling theory and other spectral analyses of light fields~\cite{chai2000plenoptic, zhang2003spectral, davis2012unstructured, mildenhall2019local} establish the sampling rates necessary to avoid aliasing, but meeting them requires specialized hardware~\cite{wilburn2005high} or a dense, time-consuming capture process~\cite{davis2012unstructured, mildenhall2019local} that is impractical for casual capture. We argue that sparsely sampling the 4D plenoptic function with commodity multi-view cameras provides a practical middle ground. By observing multiple views per exposure, we obtain parallax cues that stabilize geometry and reduce floating artifacts without requiring camera motion.

\subsection{Camera Models}
\label{sec:cameras}
We design a benchmark to measure whether using all cameras on a consumer device improves novel view synthesis over using a single camera. We model each camera $\mathbf{C}$ as a collection of sub-views $\mathbf{c}_m$, each with intrinsics $\mathbf{K}_m$ and extrinsics $(\mathbf{R}_m, \mathbf{t}_m)$. A single exposure synchronously captures all sub-views, either by triggering all cameras on a shared hardware clock or through one raw sensor readout. A sub-view is one calibrated image within that exposure, such as a physical camera's image or one decoded sub-aperture view. A monocular camera therefore contributes one viewpoint per exposure, while multi-view devices observe the same scene state from several viewpoints at once. \Cref{fig:method} summarizes the reconstruction pipeline. Given the sub-views captured in one or more exposures, we optimize 3D or 4D Gaussians~\cite{kerbl20233d, wu20244d} by simulating each camera's forward model and supervising against the captured measurements.

We evaluate four multi-view camera models that span both capture regimes, whose measurements are shown in \cref{fig:method}:
\begin{enumerate}
    \renewcommand{\labelenumi}{(\roman{enumi})}
    \item Smartphone arrays: three sub-cameras at different focal lengths (wide, ultrawide, telephoto), modeled after the iPhone 15 Pro~\cite{iphone};
    \item Stereo cameras: two horizontally separated sub-cameras;
    \item Light field cameras: a grid of sub-aperture views produced by a microlens array; and
    \item Multiplexed light field cameras: the same grid of sub-aperture views with sub-lens images that intentionally overlap on the sensor, which we describe in \cref{sec:sensor-limited-analysis}.
\end{enumerate}

To characterize the amount of parallax each camera measures through its sub-views, we define the angular baseline $\theta$ as the maximum angle between distinct sub-views, when viewed from the scene's center of gravity $\mathbf{g}$:
\begin{equation}
    \theta(\mathbf{C}, \mathbf{g}) = \max_{i,j} \cos^{-1} \left(\hat{\mathbf{v}}_i^{\top}\hat{\mathbf{v}}_j\right),
    \label{eq:angular-baseline}
\end{equation}
where $\hat{\mathbf{v}}_k = (\mathbf{t}_k-\mathbf{g})/\|\mathbf{t}_k-\mathbf{g}\|$ are unit viewing directions. For real-world scenes, we compute the center of gravity $\mathbf{g}$ as the centroid of the sparse 3D point cloud produced during pose estimation; for synthetic scenes, we use the known scene center. Angular baseline has been used previously to reason about parallax in multi-view stereo tasks~\cite{yao2018mvsnet}. The metric is scale-invariant and more meaningful than metric baseline when comparing cameras across different scene depths: a monocular camera has $\theta = 0$, and $\theta$ increases with wider baselines or closer scenes. On our real-world dataset, the median angular baselines are $5.06\degree$ (iPhone), $15.1\degree$ (stereo), and $0.74\degree$ (Lytro).

\subsection{Sensor-Limited Multi-View}
\label{sec:sensor-limited-analysis}

Under a fixed sensor budget, angular samples must be traded off with spatial resolution. A monocular camera allocates all of its pixels to one view, which maximizes per-view detail but provides no parallax within an exposure. Without parallax, geometry is unconstrained along every ray, and reconstruction from few exposures degenerates into floating artifacts and incorrect depth. A conventional light field camera divides the same sensor into an $M \times M$ grid of sub-aperture views, so each view retains only $1/M^2$ of the pixels, but the exposure now observes the scene from $M^2$ distinct directions. When exposures are scarce, we show this trade to be favorable, since the angular samples triangulate geometry that no amount of spatial resolution in a single view can recover.

\begin{figure}[t]
\centering
\includegraphics[width=\linewidth]{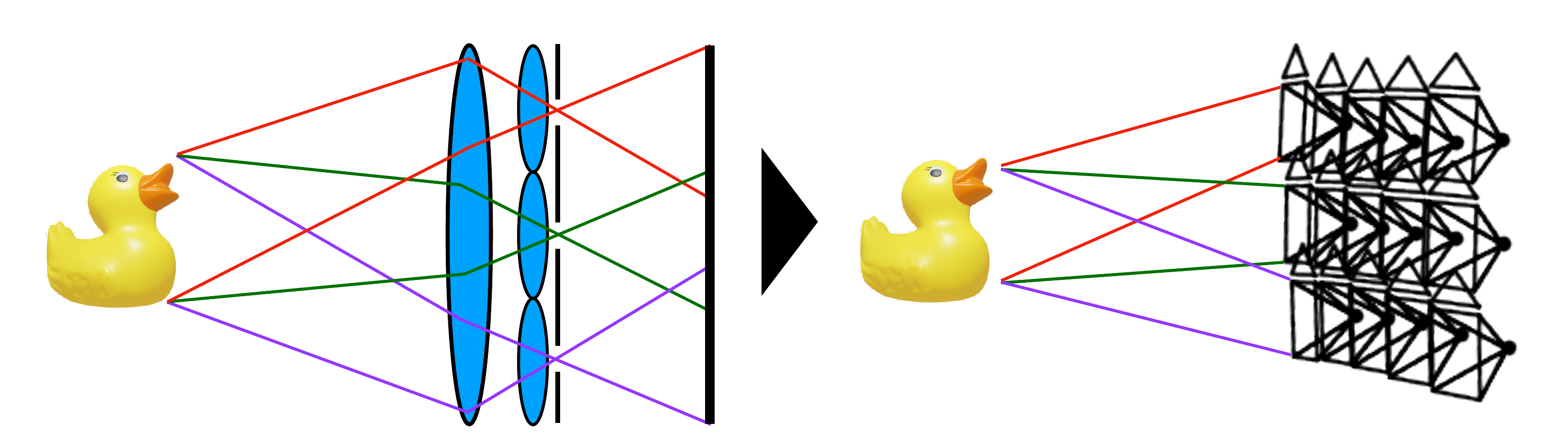}
\caption{\textbf{Multiplexed light field camera.} A lens array and aperture array behind the main lens produce sub-lens images that overlap on the sensor (\emph{left}). Each sub-view keeps a larger pixel footprint than in a non-overlapping design, and the superposed measurement is disentangled during reconstruction by simulating the forward model on rasterized Gaussians (\emph{right}).}
\label{fig:multiplexed-camera}
\end{figure}

Unlike microlens-based plenoptic cameras that partition the image sensor into non-overlapping sub-regions, we introduce a multiplexed light field camera, in which a lens array and an aperture array are placed behind the main lens so that sub-lens footprints intentionally overlap on the sensor, as shown in \cref{fig:multiplexed-camera}, exploiting compressive sensing principles to improve the angular-spatial resolution tradeoff~\cite{georgiev2006light, wilburn2005high, marwah2013compressive, liang2008programmable}. We perform a one-time calibration by masking all but one lenslet in turn, and estimating each sub-view's camera parameters and pose by capturing an image of a textured scene. We also measure each sub-view's field of view by imaging a uniform scene.

We rasterize every sub-view from its calibrated pose, apply the forward model by linearly superimposing the sub-lens images to produce the composite sensor image $\hat{y}$, and supervise against the measured image $y$. This formulation allows us to train from a single measurement. The inverse problem is harder, but each measurement carries more information per pixel, and our experiments show that the trade pays off both in simulation and on a physical prototype (\cref{sec:sensor-results}). Additional implementation details are available in the Supplementary Material.

To help disentangle the overlapping views, we modify the 3DGS training objective to include total variation regularization~\cite{rudin1992nonlinear}:
\begin{equation}
    \mathcal{L} = (1 - \lambda_1) \mathcal{L}_1 + \lambda_1 \mathcal{L}_{\text{D-SSIM}} + \lambda_2 \mathcal{L}_{\text{TV}} ,
\end{equation}
where $\mathcal{L}_{\text{TV}}$ applies 2D total variation to training images to reduce noise while preserving image edges. During each training iteration, we also choose $k$ random views facing the scene, render them using the single-lens forward model, and apply TV loss. We empirically find that this unsupervised regularization is useful for light field and multiplexed camera inputs.

\subsection{Exposure-Limited Multi-View}
\label{sec:exposure-limited-analysis}

Stereo and smartphone cameras use separate physical sensors, so every exposure yields two or three full-resolution angular samples. A moving monocular camera can match the image count of an $M$-camera device only by using $M{\times}$ more capture time. For dynamic scenes, there is no equivalent formulation, as sequential exposures observe different scene states and therefore provide no parallax at any single timestep. Synchronized sub-views are the only source of single-timestep parallax available to a casual capture rig. We therefore expect the gains in this setting to be largest for dynamic content and single-exposure capture, and to shrink as camera motion accumulates angular coverage.

We treat each exposure as a synchronized group of sub-views, rasterize each sub-view independently using its calibrated parameters, and apply the photometric loss on each sub-view independently. To isolate the effect of this multi-view from each commodity camera's photometrics, we evaluate each camera model against a monocular stream from the same camera. For dynamic scenes, we use 4D Gaussian Splatting~\cite{wu20244d}, which maintains a canonical set of 3D Gaussians and deforms them via a learned deformation network parameterized by HexPlanes~\cite{cao2023hexplane}. Because no consumer light field cameras have video capture capabilities, we perform dynamic scene evaluation on the smartphone and stereo cameras only.

%% file: sec/4_experiments.tex
\section{Experiments}
\label{sec:experiments}

We evaluate whether multi-view cameras improve novel view synthesis over their respective monocular baselines across both static and dynamic scenes. In simulation, we study how the number of exposures and angular baseline affect image and depth quality. We collect real-world data from static and dynamic scenes using consumer multi-view cameras and our multiplexed light field prototype. We show that even significantly sub-optimal angular sampling densities---as few as 2--3 simultaneous views---substantially improve reconstruction compared to monocular capture. Per-scene metrics are in the Supplementary Material.

\subsection{Multi-View Dataset}
\label{sec:multi-view-dataset}

Because no dataset compares multiple camera types on the same scenes, we collect and release two real-world datasets comprising 16 scenes (8 static, 8 dynamic), a scale consistent with established NVS evaluation datasets~\cite{mildenhall2020nerf, barron2022mipnerf, gao2022monocular}.

\subsubsection{Static scene dataset.}
We capture a dataset of eight static scenes using multiple cameras: (i) an iPhone 15 Pro with wide, ultrawide, and telephoto lenses~\cite{iphone}, (ii) an Apple Vision Pro with stereo left/right cameras~\cite{apple}, and (iii) a Lytro Illum light field camera, which produces a $13 \times 13$ grid of sub-aperture images~\cite{ng2005light}. To reduce aberrations, we use the inner $9 \times 9$ images from the Lytro, resulting in 81 sub-views per exposure. For each scene, we capture one exposure using each camera plus two casual, handheld videos using the iPhone and stereo cameras; we exclude the light field camera from video evaluation because it cannot record video. We use VGGT~\cite{wang2025vggt} to estimate camera parameters for each sub-view, followed by an additional bundle adjustment step for improved accuracy, and we initialize training from the VGGT point cloud subsampled to 100K points. The iPhone and Vision Pro cameras are hardware-synchronized via iOS's AVMultiCamSession API, which locks all active camera streams to the same hardware clock, ensuring sub-frame temporal alignment across all sensors.

\begin{figure}[t]
\centering
\includegraphics[width=\linewidth]{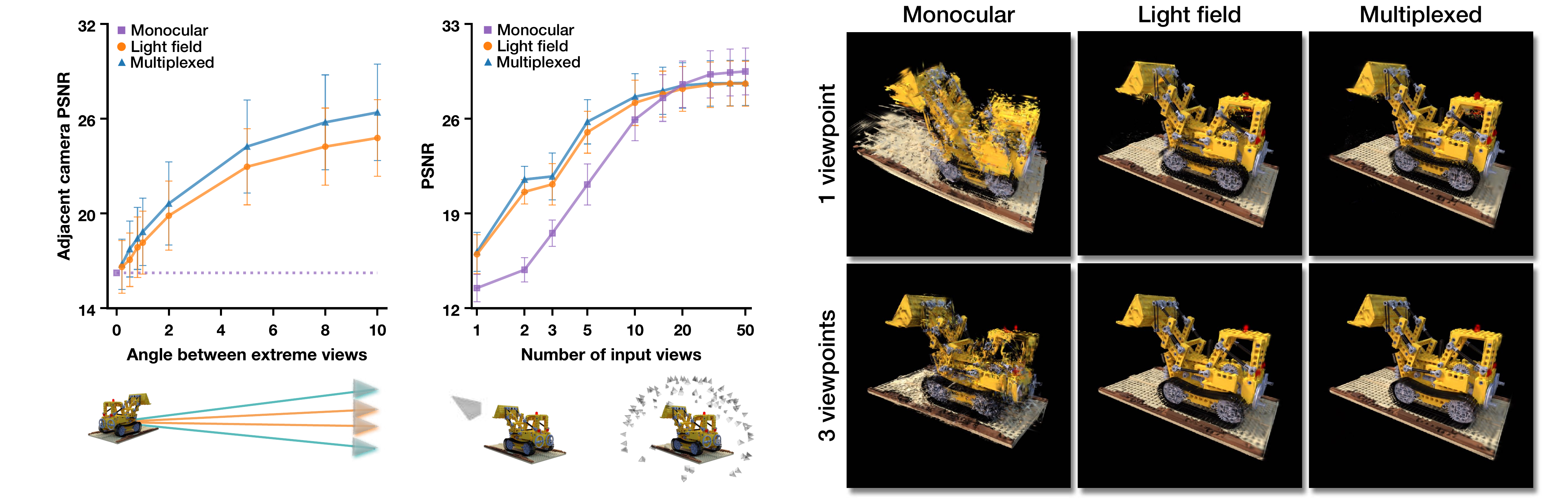}
\caption{\textbf{Sensor-limited synthetic analysis.} With a single exposure (\emph{left}), both light field designs outperform the monocular camera, with the benefit growing as the angular baseline increases. As the number of exposures increases (\emph{center}), all cameras converge in performance. In the reconstructions (\emph{right}), angular sampling suppresses artifacts despite lower per-view resolution.}
\label{fig:synthetic-lightfield}
\end{figure}

\begin{figure}[t]
\centering
\includegraphics[width=\linewidth]{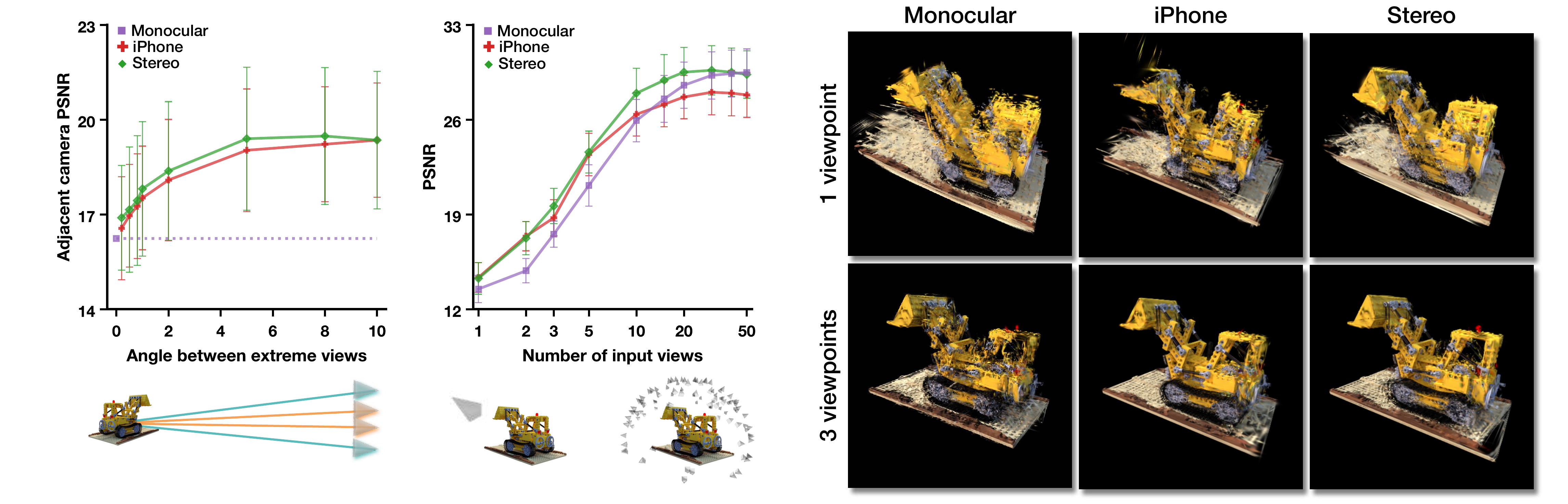}
\caption{\textbf{Exposure-limited synthetic analysis.} With a single exposure (\emph{left}), the benefits of multi-view cameras grow as the angular baseline increases. As the number of exposures increases (\emph{center}), all cameras converge in performance.}
\label{fig:synthetic-exposure}
\end{figure}

\subsubsection{Dynamic scene dataset.}
To evaluate light field sampling's effect on dynamic reconstruction, we capture multi-view videos using the three cameras on the iPhone and the two cameras on the Vision Pro across eight real-world dynamic scenes. We time-subsample these videos to 10\,FPS for the iPhone and 15\,FPS for the stereo camera to accommodate GPU memory. We estimate monocular pose using VGGT, generate frame-level poses on one video per camera rig using MegaSaM~\cite{li2025megasam}, and propagate them to the other sub-views, since the relative poses between sub-views are constant.

\subsection{Synthetic Experiments}
\label{sec:synthetic-results}

We evaluate on all scenes from the NeRF Synthetic dataset~\cite{mildenhall2020nerf}. We render synthetic training views for each camera model using the original Blender files. The sensor-limited light field camera captures a $4 \times 4$ grid of sub-aperture views, and the multiplexed camera captures the same grid with overlapping images at higher spatial resolution per sub-view. The exposure-limited models render each sub-view at full resolution from its own sensor. For the stereo camera, we generate two views per exposure with a horizontal baseline. The iPhone model produces three views at different focal lengths simulating the wide, ultrawide, and telephoto lenses~\cite{iphone}.

\input{tables/synthetic-summary}

\input{tables/synthetic-depth.tex}

\begin{figure}[t]
\centering
\includegraphics[width=\linewidth]{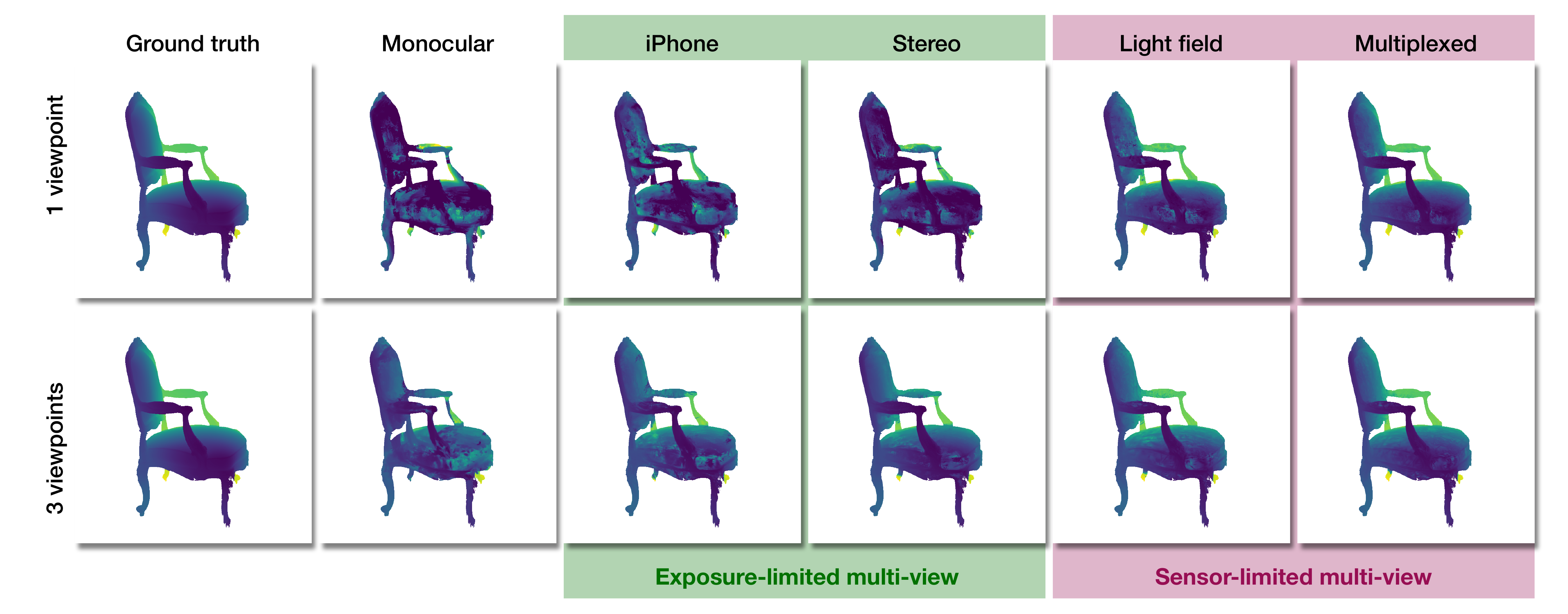}
\caption{\textbf{Predicted depth on the \emph{chair} scene.}
Ground truth and predicted depth for one exposure (\emph{top}) and three exposures (\emph{bottom}). The multi-view models recover more coherent depth than the monocular reconstruction.}
\label{fig:synthetic-depth}
\end{figure}

We conduct two experiments that vary angular baseline and exposure count, respectively. For the angular baseline study, we select $N \in \{1, 3\}$ exposures using max-min dispersion to ensure pose diversity and sweep the angular baseline from $\theta = 0.2\degree$ to $10\degree$. For the exposure count study, we evaluate how NVS quality scales with $N \in \{1, \ldots, 50\}$ at a fixed $\theta = 10\degree$. Because a single exposure cannot observe the full scene, single-exposure runs are evaluated on the six test views nearest the training pose. The multi-exposure runs use the full 200-image test set. We initialize Gaussians by sampling uniformly at random within a sphere containing the target object rather than a cube, which reduces floating artifacts, as we show in the Supplementary Material.

To better match real camera characteristics, we simulate noisy sensor measurements using Gaussian-Poisson noise. Following~\cite{brooks2019unprocessing}, we model noisy observations as $y \sim \mathcal{N}(x, \sigma^2(x))$, where $\sigma^2(x) = \lambda_{\text{read}} + \lambda_{\text{shot}}\,x$ combines signal-independent read noise and signal-dependent shot noise. We calibrate $\lambda_{\text{read}}$ and $\lambda_{\text{shot}}$ on images from a Canon R10 camera, and apply 14-bit quantization.

\cref{tab:synthetic-summary} (left) and \cref{fig:synthetic-lightfield} show that both light field cameras outperform monocular capture under the same sensor budget when exposures are scarce. This demonstrates that angular diversity outweighs spatial resolution when views are limited. Multiplexing further improves performance, even when the additional spatial resolution is achieved through overlapping. As the number of exposures grows, all designs converge in performance. In the exposure-limited setting, the iPhone and stereo models likewise outperform their monocular baselines at matched exposure time, as shown in \cref{tab:synthetic-summary} (right) and \cref{fig:synthetic-exposure}.

To evaluate geometric fidelity, we compare predicted depth against ground truth on the NeRF Synthetic scenes. We use the same reconstructions, camera models, and capture configurations as in the image-based experiments, with $N\in\{1,3\}$ exposures and angular baseline $\theta=10\degree$. For each held-out camera, we obtain ground truth depth by rendering depth from the original Blender scene. We report absolute relative error and root mean squared error on the same test views as the image-based experiments in \cref{tab:synthetic-depth}. All four light field sampling cameras reduce geometric error relative to monocular capture at both exposure counts, with the largest improvement provided by the light field and multiplexed light field models. The predicted depth maps in \cref{fig:synthetic-depth} show the same trend qualitatively.

\begin{figure}[!t]
\centering
\includegraphics[width=\linewidth]{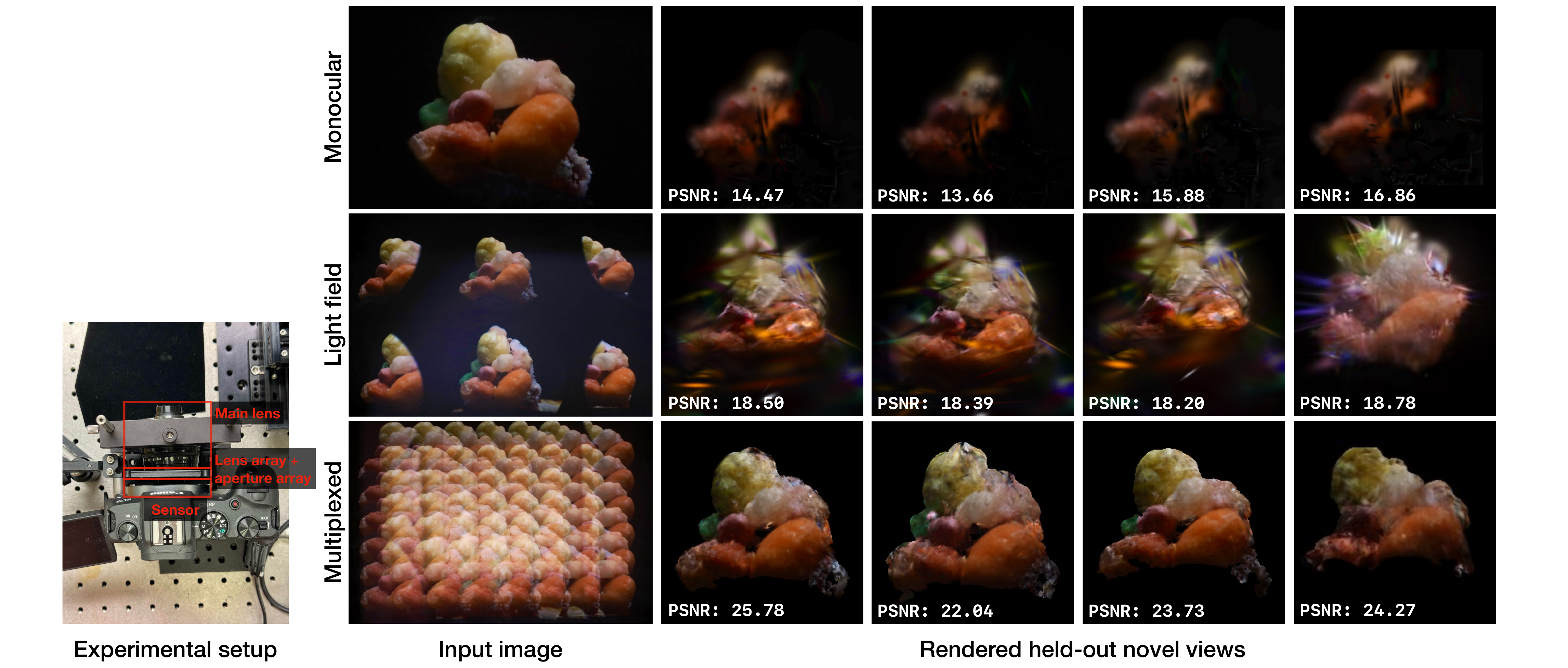}
\caption{\textbf{Multiplexed light field.} We capture a scene with our prototype (\emph{left}) using monocular, light field, and multiplexed configurations (\emph{middle}), and render held-out novel views from each single-exposure reconstruction (\emph{right}). The multiplexed camera recovers geometry that monocular capture cannot, and retains detail that the non-overlapping light field loses to reduced per-view resolution.}
\label{fig:multiplexed-result}
\end{figure}

\subsection{Real-World Static Scenes}
\label{sec:sensor-results}

We reconstruct the static scenes in our dataset from a single exposure per camera. We compare each device against its own monocular baseline on held-out evaluation views. Following DyCheck~\cite{gao2022monocular}, we report co-visibility masked metrics (mPSNR, mSSIM, mLPIPS) to evaluate only test pixels that were adequately observed during training, preventing penalization for failing to hallucinate unobserved regions. Alongside 3DGS, we evaluate the sparse-view baselines SparseGS~\cite{xiong2025sparsegs} and FSGS~\cite{zhu2024fsgs}, which incorporate additional learned regularizers. Hyperparameters were selected through a standard grid search with the same tuning budget for each camera model and method, and the best configuration per camera model was fixed across all scenes.

\input{tables/static-real-summary}

\begin{figure}[!t]
\centering
\includegraphics[width=\linewidth]{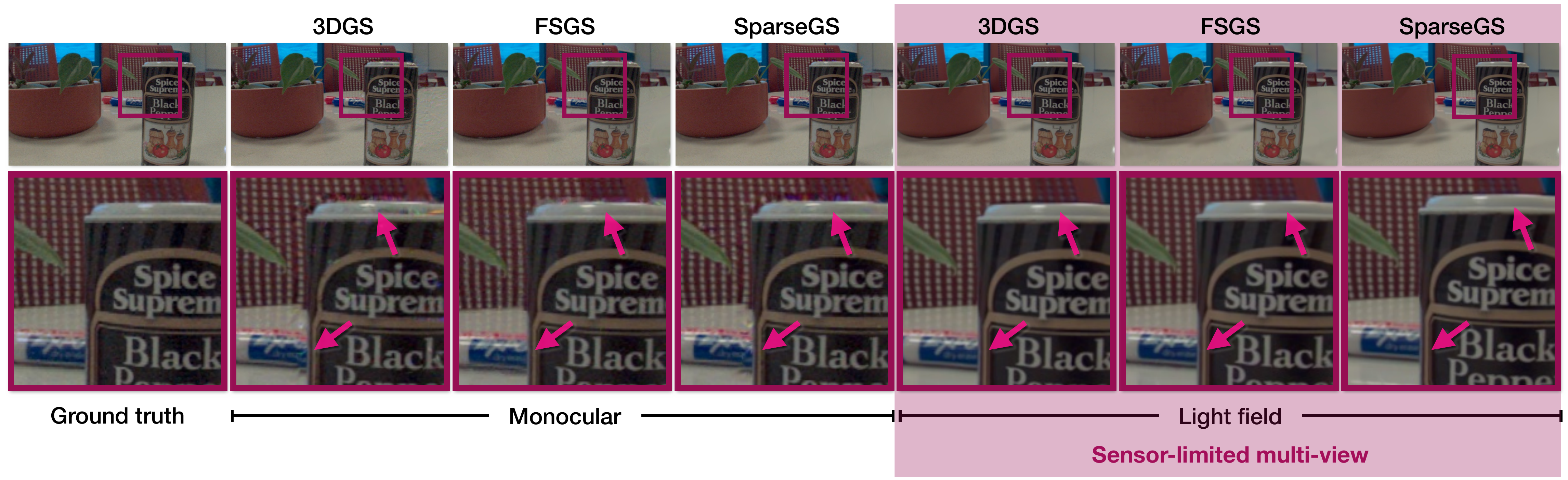}
\caption{\textbf{Real, single-exposure reconstruction (sensor-limited).} We reconstruct a scene from one Lytro exposure using either one monocular view or 81 smaller sub-views. The additional angular samples from the same exposure recover structure that the center view leaves underdetermined, and this benefit complements sparse-view methods that incorporate learned priors.}
\label{fig:static-lightfield}
\end{figure}

\subsubsection{Sensor-limited.}
We evaluate the light field tradeoff using the Lytro Illum, comparing the central sub-aperture view alone against all 81 sub-views. We downsample the other sub-views such that together they have the same number of pixels as the monocular image. As shown in \cref{tab:static-real-summary}, using all sub-views improves mPSNR and mSSIM within every reconstruction method. However, because the light field sub-views have low spatial resolution compared to the monocular image, mLPIPS is sensitive to the texture softness even where geometry improves. \Cref{fig:static-lightfield} shows qualitative results where the monocular reconstructions break down at occlusion boundaries that the full light field is able to resolve.

To validate the multiplexed design, we construct a physical prototype consisting of a lens array and aperture array mounted behind the main lens of a mirrorless camera (\cref{fig:multiplexed-result}). Additional implementation details are available in the Supplementary Material. We capture a static scene and compare three configurations from a single exposure: a monocular image, a conventional light field input formed from non-overlapping sub-lens captures, and a multiplexed exposure with overlapping sub-images. As shown in \cref{fig:multiplexed-result}, the multiplexed configuration recovers scene geometry where the monocular reconstruction collapses, and retains detail that the non-overlapping light field loses.

\begin{figure}[!tbp]
\centering
\includegraphics[width=\linewidth]{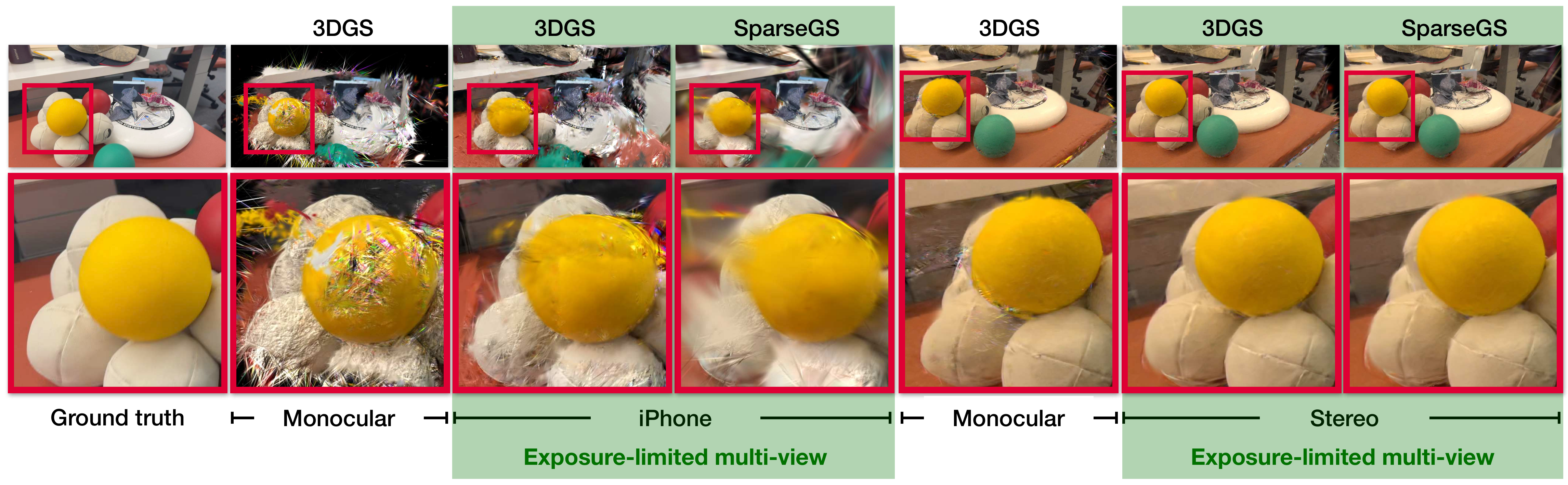}
\caption{\textbf{Real single-exposure reconstruction (exposure-limited).} Synchronized views sharpen geometry for both the iPhone and stereo cameras.}
\label{fig:static-exposure}
\end{figure}

\input{tables/exposure-limited-casual}

\subsubsection{Exposure-limited.}

We compare each device against its own monocular baseline: the wide camera on the iPhone, and the right camera on the stereo camera. As shown in \cref{tab:static-real-summary}, synchronized views improve all metrics compared to the monocular baseline. The additional measurement views complement the learned regularizers in the sparse-view methods. \Cref{fig:static-exposure} shows qualitative results.

\begin{figure}[!tbp]
\centering
\includegraphics[width=\linewidth]{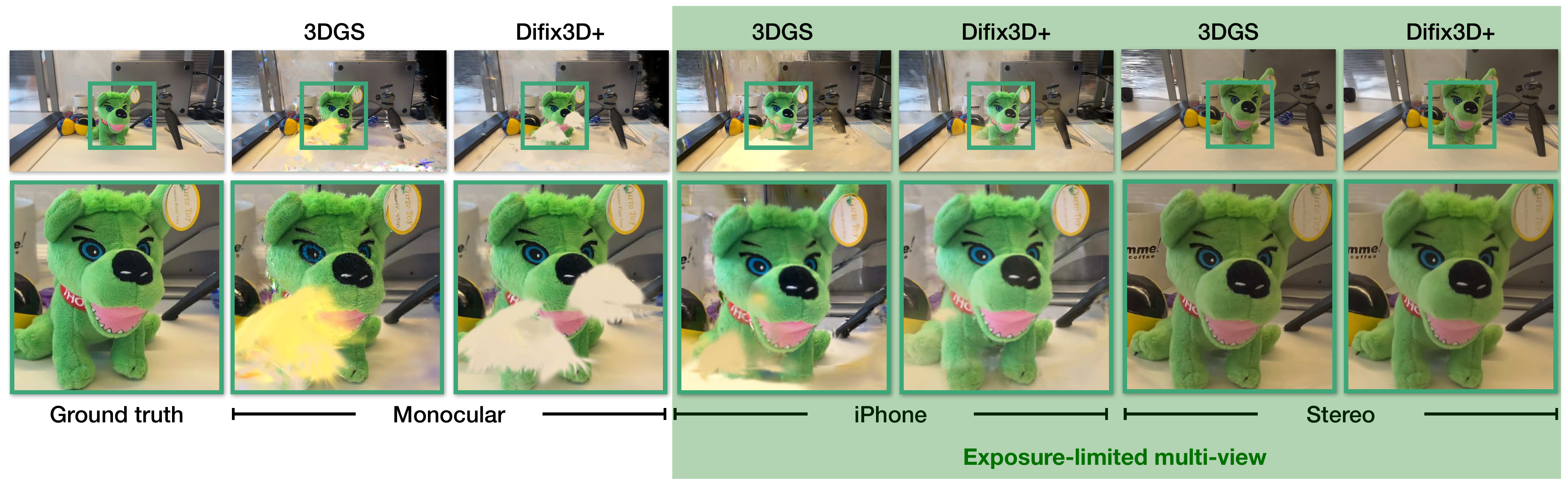}
\caption{\textbf{Casual video reconstruction.} We reconstruct a handheld capture from the monocular stream and from all synchronized views, with and without the Difix3D+ prior. Camera motion already supplies angular diversity, so multi-view gains are modest, and either added views or the prior can repair the monocular ghosting.}
\label{fig:casual}
\end{figure}

We also evaluate both devices on casual handheld video of the same scenes, comparing each multi-view capture against its monocular baseline. We show additional experiments where we augment training with Difix3D+~\cite{wu2025difix3d}, a recent method that trains 3DGS with a diffusion-based artifact correction step that cleans rendered images and injects them back into training. Results are shown in \cref{tab:exposure-limited-casual} and \cref{fig:casual}. Both multi-view cameras show improvements over their monocular counterparts. The gains are smaller than in the single-exposure and dynamic settings, which is expected: the value of multi-camera input is largest in regimes where monocular capture is most constrained, and smallest when casual video already supplies sufficient angular diversity.

\begin{figure}[!t]
\centering
\includegraphics[width=\linewidth]{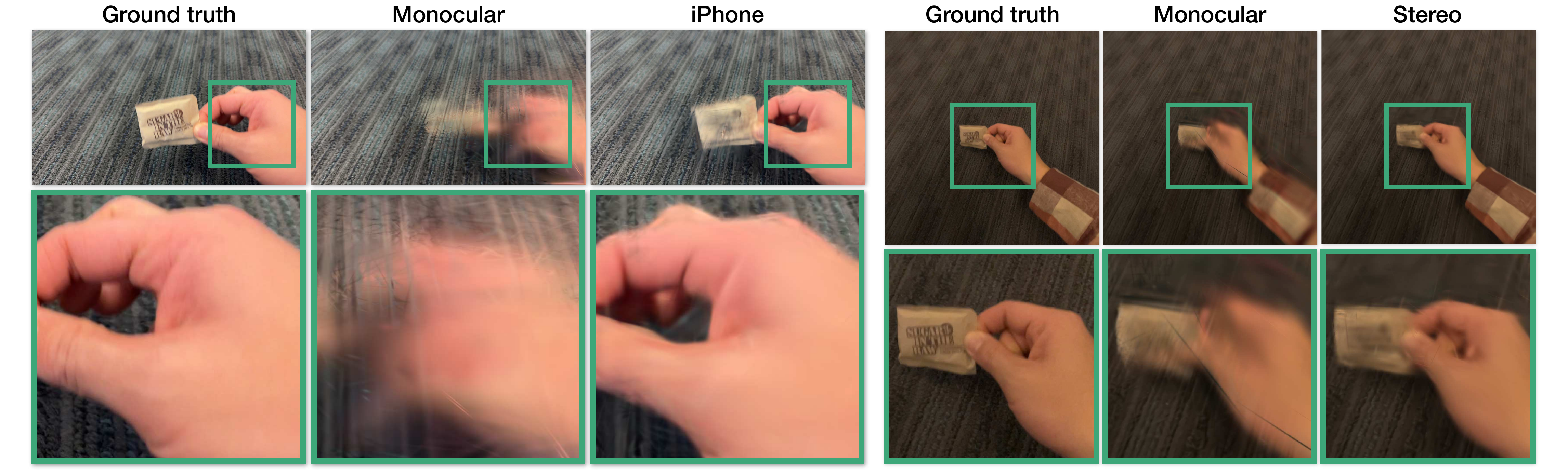}
\caption{\textbf{Dynamic reconstruction comparison.} Because it provides no effective multi-view cues, a stationary monocular camera cannot recover scene motion, and the moving hand and object smear across time. Using all cameras on the iPhone (\emph{left}) or stereo device (\emph{right}) provides the angular diversity needed for 4DGS to reconstruct dynamic content.}
\label{fig:dynamic}
\end{figure}

\subsection{Real-World Dynamic Scenes}
\label{sec:dynamic-results}

We reconstruct each dynamic scene using 4DGS, comparing each device's multi-view stream against its own monocular stream. Following 4DGS, we evaluate held-out frames from the training sequence (\cref{tab:exposure-limited-dynamic}). The gains are largest in this setting because sequential monocular exposures observe different scene states. As shown in \cref{fig:dynamic}, monocular input smears motion---a limitation also noted in the 4DGS paper~\cite{wu20244d}---whereas synchronized multi-camera input recovers geometry and dynamics.

\input{tables/exposure-limited-dynamic}

%% file: tables/synthetic-summary.tex
\begin{table}[!htbp]
\caption{\textbf{Results on synthetic scenes.} Average metrics across all scenes in the Blender dataset~\cite{mildenhall2020nerf} (angle=10$^{\circ}$). 1-view methods report adjacent camera metrics, whereas 3-view methods report full test camera metrics.}
\label{tab:synthetic-summary}
\centering
\scriptsize
\setlength{\tabcolsep}{2.5pt}
\renewcommand{\arraystretch}{1.0}
\begin{tabular}[t]{clccc}
\multicolumn{5}{c}{\textit{Sensor-limited}} \\
\toprule
Exp. & Camera & PSNR$\uparrow$ & SSIM$\uparrow$ & LPIPS$\downarrow$ \\
\midrule
\multirow{3}{*}{1} & Monocular & 16.25 & 0.696 & 0.257 \\
 & Light field & 24.78 & 0.869 & 0.122 \\
 & Multiplexed & \textbf{26.39} & \textbf{0.891} & \textbf{0.098} \\
\midrule
\multirow{3}{*}{3} & Monocular & 17.53 & 0.763 & 0.210 \\
 & Light field & 22.94 & 0.848 & 0.138 \\
 & Multiplexed & \textbf{23.85} & \textbf{0.866} & \textbf{0.121} \\
\bottomrule
\end{tabular}
\hspace{1.5em}
\begin{tabular}[t]{clccc}
\multicolumn{5}{c}{\textit{Exposure-limited}} \\
\toprule
Exp. & Camera & PSNR$\uparrow$ & SSIM$\uparrow$ & LPIPS$\downarrow$ \\
\midrule
\multirow{3}{*}{1} & Monocular & 16.25 & 0.696 & 0.257 \\
 & iPhone & \textbf{19.35} & 0.759 & \textbf{0.203} \\
 & Stereo & \textbf{19.35} & \textbf{0.768} & \textbf{0.203} \\
\midrule
\multirow{3}{*}{3} & Monocular & 17.53 & 0.763 & 0.210 \\
 & iPhone & 19.92 & 0.800 & 0.173 \\
 & Stereo & \textbf{20.69} & \textbf{0.827} & \textbf{0.151} \\
\bottomrule
\end{tabular}
\end{table}

%% file: tables/synthetic-depth.tex
\begin{table}[!htbp]
\caption{\textbf{Synthetic scene depth accuracy.} Mean geometric (depth) errors over the NeRF Synthetic dataset at $\theta=10\degree$.}
\label{tab:synthetic-depth}
\centering
\small
\begingroup
\setlength{\tabcolsep}{3pt}
\begin{tabular}{l@{}p{1.75em}@{}cc@{}p{2.25em}@{}cc}
\toprule
& & \multicolumn{2}{c}{1 exposure}
& & \multicolumn{2}{c}{3 exposures} \\
\cmidrule(lr){3-4}\cmidrule(lr){6-7}
Camera model
& & AbsRel$\downarrow$ & RMSE$\downarrow$
& & AbsRel$\downarrow$ & RMSE$\downarrow$ \\
\midrule
Monocular
& & 0.0939 & 0.4588
& & 0.0557 & 0.3001 \\
\addlinespace
\multicolumn{7}{l}{\textit{Exposure-limited multi-view}} \\
iPhone
& & 0.0600 & 0.3292
& & 0.0431 & 0.2541 \\
Stereo
& & \textbf{0.0526} & \textbf{0.3038}
& & \textbf{0.0314} & \textbf{0.1960} \\
\addlinespace
\multicolumn{7}{l}{\textit{Sensor-limited multi-view}} \\
Light field
& & 0.0212 & 0.1347
& & 0.0206 & 0.1200 \\
Multiplexed
& & \textbf{0.0181} & \textbf{0.1135}
& & \textbf{0.0187} & \textbf{0.1098} \\
\bottomrule
\end{tabular}
\endgroup
\end{table}

%% file: tables/static-real-summary.tex
\begin{table}[!htbp]
\caption{\textbf{Results on real-world single-exposure static scenes.} We compare the angular sampling camera against its monocular baseline in both sensor-limited (\emph{left}) and exposure-limited (\emph{right}) regimes.}
\label{tab:static-real-summary}
\centering
\scriptsize
\setlength{\tabcolsep}{2pt}
\renewcommand{\arraystretch}{1.0}
\resizebox{\linewidth}{!}{%
\begin{tabular}[t]{lccc}
\multicolumn{4}{c}{\textbf{Sensor-limited}} \\
\toprule
Method & mPSNR$\uparrow$ & mSSIM$\uparrow$ & mLPIPS$\downarrow$ \\
\midrule
\multicolumn{4}{l}{\textit{Light field capture}} \\
Mono. + 3DGS & 26.24 & 0.656 & 0.243 \\
Mono. + FSGS & 27.35 & 0.703 & 0.250 \\
Mono. + SparseGS & 27.75 & 0.712 & \textbf{0.231} \\
LF + 3DGS & \textbf{30.50} & \textbf{0.818} & 0.233 \\
LF + FSGS & 29.72 & 0.793 & 0.287 \\
LF + SparseGS & 30.37 & 0.815 & 0.244 \\
\bottomrule
\end{tabular}
\hspace{4pt}
\begin{tabular}[t]{lccc}
\multicolumn{4}{c}{\textbf{Exposure-limited}} \\
\toprule
Method & mPSNR$\uparrow$ & mSSIM$\uparrow$ & mLPIPS$\downarrow$ \\
\midrule
\multicolumn{4}{l}{\textit{iPhone capture}} \\
Mono. + 3DGS & 15.69 & 0.759 & 0.478 \\
Mono. + FSGS & 17.41 & 0.824 & 0.388 \\
Mono. + SparseGS & 19.16 & 0.840 & 0.354 \\
iPhone + 3DGS & 18.08 & 0.811 & 0.432 \\
iPhone + FSGS & 19.55 & \textbf{0.859} & 0.364 \\
iPhone + SparseGS & \textbf{19.81} & 0.855 & \textbf{0.338} \\
\midrule
\multicolumn{4}{l}{\textit{Stereo capture}} \\
Mono. + 3DGS & 17.57 & 0.862 & 0.464 \\
Mono. + FSGS & 18.29 & 0.885 & 0.451 \\
Mono. + SparseGS & 17.73 & 0.868 & 0.451 \\
Stereo + 3DGS & 20.89 & 0.901 & 0.419 \\
Stereo + FSGS & \textbf{21.57} & \textbf{0.923} & \textbf{0.399} \\
Stereo + SparseGS & 21.15 & 0.910 & 0.403 \\
\bottomrule
\end{tabular}}
\end{table}

%% file: tables/exposure-limited-casual.tex
\begin{table}[!ht]
\caption{\textbf{Results on real-world casual scenes.} Each device is evaluated against its own monocular stream, with and without the \textsc{Difix3D+}~\cite{wu2025difix3d} prior.}
\label{tab:exposure-limited-casual}
\centering
\footnotesize
\renewcommand{\arraystretch}{1.0}
\begin{tabular}{lccc}
\toprule
Camera & mPSNR$\uparrow$ & mSSIM$\uparrow$ & mLPIPS$\downarrow$ \\
\midrule
\multicolumn{4}{l}{\textit{iPhone capture}} \\
Monocular & 21.75 & 0.822 & 0.234 \\
Monocular w/ prior & \textbf{23.67} & \textbf{0.866} & \textbf{0.174} \\
iPhone & 22.12 & 0.810 & 0.251 \\
iPhone w/ prior & 23.25 & 0.841 & 0.198 \\
\midrule
\multicolumn{4}{l}{\textit{Stereo capture}} \\
Monocular & 27.89 & 0.911 & 0.205 \\
Monocular w/ prior & 28.93 & \textbf{0.937} & \textbf{0.189} \\
Stereo & 28.44 & 0.922 & 0.191 \\
Stereo w/ prior & \textbf{28.97} & \textbf{0.937} & \textbf{0.189} \\
\bottomrule
\end{tabular}
\end{table}

%% file: tables/exposure-limited-dynamic.tex
\begin{table}[!htbp]
\caption{\textbf{Results on real-world, dynamic scenes.} We compare iPhone and stereo cameras to their monocular baselines on dynamic 4DGS reconstruction.}
\label{tab:exposure-limited-dynamic}
\centering
\footnotesize
\renewcommand{\arraystretch}{1.0}
\begin{tabular}{lccc}
\toprule
Camera & PSNR$\uparrow$ & SSIM$\uparrow$ & LPIPS$\downarrow$ \\
\midrule
\multicolumn{4}{l}{\textit{iPhone capture}} \\
Monocular & 24.51 & 0.879 & 0.280 \\
iPhone & \textbf{31.71} & \textbf{0.920} & \textbf{0.228} \\
\midrule
\multicolumn{4}{l}{\textit{Stereo capture}} \\
Monocular & 29.17 & 0.939 & 0.320 \\
Stereo & \textbf{34.48} & \textbf{0.955} & \textbf{0.302} \\
\bottomrule
\end{tabular}
\end{table}

%% file: sec/5_discussion.tex
\section{Discussion}
\label{sec:discussion}

Despite the wide availability of multi-view consumer cameras, most methods for 3D and 4D radiance field reconstruction consider only monocular inputs. In this paper, we show that across synthetic and real-world experiments, synchronized multi-view capture improves single-exposure and dynamic reconstruction and complements learned sparse-view priors. Under a fixed sensor budget, angular samples are more valuable than spatial resolution when exposures are scarce, while multiplexing recovers part of the lost resolution. The gains depend on angular coverage and scene depth: denser sampling, wider baselines, and nearby objects provide stronger parallax. Benefits may therefore diminish for distant content in large, unbounded scenes where disparity is minimal at more distant depths.

Multi-view consumer cameras represent a major underexploited resource for novel view synthesis, particularly in scenarios with limited views and dynamic content. We urge the community to record all available camera streams in future datasets, and we hope our results inform future camera designs built jointly with reconstruction algorithms.

%% file: tables/static-synthetic-per-scene.tex
\begin{table}[htbp]
\centering
\scriptsize
\renewcommand{\arraystretch}{0.95}

\textbf{PSNR$\uparrow$}\\[2pt]
\resizebox{\linewidth}{!}{%
\begin{tabular}{clcccccccccc}
\toprule
Views & Method & Chair & Drums & Ficus & Hotdog & Lego & Materials & Mic & Ship & Avg & Std \\
\midrule
\multirow{5}{*}{1} & Monocular & 16.28 & 13.86 & 20.09 & 12.11 & 15.41 & 13.57 & 20.53 & 18.12 & 16.25 & 2.90 \\
 & iPhone & 22.65 & 17.33 & 22.79 & 13.62 & 20.85 & 15.58 & 23.24 & 18.73 & 19.35 & 3.38 \\
 & Stereo & 20.26 & 15.93 & 24.12 & 14.87 & 17.20 & 15.31 & 26.79 & 20.33 & 19.35 & 4.07 \\
 & Light field & 27.16 & 19.81 & 31.45 & 17.39 & 24.64 & 23.83 & 30.41 & 23.53 & 24.78 & 4.52 \\
 & Multiplexed & 29.78 & 20.08 & 33.77 & 16.86 & 28.38 & 24.18 & 33.59 & 24.53 & 26.39 & 5.71 \\
\midrule
\multirow{5}{*}{3} & Monocular & 15.38 & 16.22 & 19.41 & 16.83 & 16.19 & 16.13 & 21.25 & 18.83 & 17.53 & 1.93 \\
 & iPhone & 18.69 & 17.73 & 21.50 & 20.25 & 17.83 & 18.40 & 25.28 & 19.64 & 19.92 & 2.36 \\
 & Stereo & 19.27 & 17.99 & 22.69 & 21.15 & 18.83 & 18.56 & 25.95 & 21.11 & 20.69 & 2.48 \\
 & Light field & 22.60 & 19.35 & 24.75 & 25.46 & 22.01 & 20.49 & 26.48 & 22.38 & 22.94 & 2.30 \\
 & Multiplexed & 23.75 & 19.77 & 26.83 & 25.59 & 23.34 & 20.79 & 27.22 & 23.51 & 23.85 & 2.49 \\
\bottomrule
\end{tabular}}
\vspace{0.3em}

\textbf{SSIM$\uparrow$}\\[2pt]
\resizebox{\linewidth}{!}{%
\begin{tabular}{clcccccccccc}
\toprule
Views & Method & Chair & Drums & Ficus & Hotdog & Lego & Materials & Mic & Ship & Avg & Std \\
\midrule
\multirow{5}{*}{1} & Monocular & 0.778 & 0.649 & 0.844 & 0.640 & 0.680 & 0.543 & 0.849 & 0.585 & 0.696 & 0.108 \\
 & iPhone & 0.878 & 0.761 & 0.897 & 0.664 & 0.810 & 0.588 & 0.886 & 0.585 & 0.759 & 0.123 \\
 & Stereo & 0.852 & 0.711 & 0.922 & 0.706 & 0.747 & 0.608 & 0.941 & 0.653 & 0.768 & 0.116 \\
 & Light field & 0.918 & 0.831 & 0.970 & 0.815 & 0.876 & 0.823 & 0.966 & 0.755 & 0.869 & 0.072 \\
 & Multiplexed & 0.950 & 0.846 & 0.983 & 0.802 & 0.937 & 0.841 & 0.983 & 0.782 & 0.891 & 0.077 \\
\midrule
\multirow{5}{*}{3} & Monocular & 0.790 & 0.731 & 0.838 & 0.810 & 0.718 & 0.697 & 0.868 & 0.650 & 0.763 & 0.070 \\
 & iPhone & 0.836 & 0.785 & 0.858 & 0.851 & 0.743 & 0.734 & 0.919 & 0.673 & 0.800 & 0.075 \\
 & Stereo & 0.859 & 0.804 & 0.889 & 0.865 & 0.782 & 0.776 & 0.929 & 0.708 & 0.827 & 0.067 \\
 & Light field & 0.872 & 0.829 & 0.907 & 0.894 & 0.806 & 0.814 & 0.929 & 0.734 & 0.848 & 0.060 \\
 & Multiplexed & 0.891 & 0.845 & 0.933 & 0.905 & 0.831 & 0.824 & 0.938 & 0.764 & 0.866 & 0.056 \\
\bottomrule
\end{tabular}}
\vspace{0.3em}

\textbf{LPIPS$\downarrow$}\\[2pt]
\resizebox{\linewidth}{!}{%
\begin{tabular}{clcccccccccc}
\toprule
Views & Method & Chair & Drums & Ficus & Hotdog & Lego & Materials & Mic & Ship & Avg & Std \\
\midrule
\multirow{5}{*}{1} & Monocular & 0.167 & 0.309 & 0.114 & 0.369 & 0.254 & 0.365 & 0.121 & 0.356 & 0.257 & 0.102 \\
 & iPhone & 0.076 & 0.210 & 0.082 & 0.317 & 0.142 & 0.340 & 0.103 & 0.355 & 0.203 & 0.111 \\
 & Stereo & 0.106 & 0.261 & 0.062 & 0.292 & 0.212 & 0.329 & 0.047 & 0.313 & 0.203 & 0.108 \\
 & Light field & 0.070 & 0.159 & 0.031 & 0.177 & 0.120 & 0.149 & 0.041 & 0.227 & 0.122 & 0.065 \\
 & Multiplexed & 0.040 & 0.138 & 0.015 & 0.179 & 0.063 & 0.126 & 0.016 & 0.203 & 0.098 & 0.069 \\
\midrule
\multirow{5}{*}{3} & Monocular & 0.178 & 0.223 & 0.135 & 0.204 & 0.250 & 0.262 & 0.126 & 0.306 & 0.210 & 0.058 \\
 & iPhone & 0.133 & 0.173 & 0.122 & 0.162 & 0.218 & 0.224 & 0.071 & 0.284 & 0.173 & 0.063 \\
 & Stereo & 0.113 & 0.158 & 0.092 & 0.152 & 0.188 & 0.194 & 0.062 & 0.251 & 0.151 & 0.057 \\
 & Light field & 0.105 & 0.144 & 0.084 & 0.127 & 0.173 & 0.170 & 0.071 & 0.234 & 0.138 & 0.050 \\
 & Multiplexed & 0.091 & 0.122 & 0.055 & 0.117 & 0.153 & 0.155 & 0.058 & 0.219 & 0.121 & 0.051 \\
\bottomrule
\end{tabular}}

\caption{Per-scene metrics across all scenes in the NeRF synthetic Blender dataset~\cite{mildenhall2020nerf} (angle=10$^{\circ}$). 1-exposure methods report adjacent camera metrics; 3-exposure methods report full test camera metrics.}
\label{tab:static-synthetic-per-scene}
\end{table}

%% file: tables/static-real-per-scene.tex
\begin{table}[htbp]
\centering
\scriptsize
\renewcommand{\arraystretch}{0.84}
\captionsetup{skip=1pt}

\textbf{mPSNR$\uparrow$}\\[1pt]
\resizebox{\linewidth}{!}{%
\begin{tabular}{lcccccccccc}
\toprule
Method & Action Figure & Ball & Chicken & Dog & Espresso & Optics & Salt Pepper & Shelf & Avg & Std \\
\midrule
\multicolumn{11}{l}{\textit{Light field capture}} \\
Monocular + 3DGS & 24.82 & 27.50 & 25.99 & 26.90 & 28.83 & 20.87 & 27.16 & 27.81 & 26.23 & 2.32 \\
Monocular + FSGS & 25.24 & 29.46 & 28.14 & 28.28 & 28.83 & 21.80 & 28.04 & 28.98 & 27.35 & 2.41 \\
Monocular + SparseGS & 24.13 & 29.80 & 28.78 & 29.01 & 29.81 & 21.81 & 29.12 & 29.57 & 27.75 & 2.84 \\
Light field + 3DGS & 30.14 & 31.64 & 30.27 & 30.96 & 32.16 & 26.35 & 30.91 & 31.56 & 30.50 & 1.70 \\
Light field + FSGS & 28.09 & 31.25 & 30.48 & 30.36 & 32.09 & 22.87 & 31.06 & 31.59 & 29.72 & 2.82 \\
Light field + SparseGS & 29.86 & 31.56 & 30.61 & 30.97 & 32.13 & 25.27 & 31.01 & 31.54 & 30.37 & 2.03 \\
\midrule
\multicolumn{11}{l}{\textit{iPhone capture}} \\
Monocular + 3DGS & 19.19 & 14.16 & 12.18 & 15.30 & 22.01 & 16.32 & 15.99 & 10.38 & 15.69 & 3.46 \\
Monocular + FSGS & 23.38 & 15.56 & 13.68 & 16.76 & 20.54 & 16.97 & 16.57 & 15.81 & 17.41 & 2.89 \\
Monocular + SparseGS & 23.26 & 15.42 & 12.89 & 18.41 & 21.46 & 15.75 & 16.78 & 29.28 & 19.16 & 4.95 \\
iPhone + 3DGS & 21.53 & 15.39 & 14.64 & 17.54 & 19.61 & 17.84 & 16.30 & 21.78 & 18.08 & 2.51 \\
iPhone + FSGS & 24.65 & 15.89 & 14.36 & 18.25 & 22.31 & 18.03 & 17.05 & 25.88 & 19.55 & 3.94 \\
iPhone + SparseGS & 24.01 & 16.55 & 15.00 & 18.51 & 22.07 & 16.47 & 17.15 & 28.70 & 19.81 & 4.41 \\
\midrule
\multicolumn{11}{l}{\textit{Stereo capture}} \\
Monocular + 3DGS & 10.26 & 20.17 & 19.26 & 26.95 & 20.66 & 11.68 & 14.33 & 17.23 & 17.57 & 5.07 \\
Monocular + FSGS & 11.71 & 19.40 & 19.98 & 26.15 & 22.94 & 12.54 & 14.07 & 19.53 & 18.29 & 4.79 \\
Monocular + SparseGS & 10.81 & 19.93 & 19.54 & 27.68 & 20.77 & 11.92 & 14.26 & 16.89 & 17.73 & 5.13 \\
Stereo + 3DGS & 16.37 & 24.21 & 20.58 & 28.10 & 26.24 & 17.24 & 18.36 & 15.99 & 20.89 & 4.41 \\
Stereo + FSGS & 14.77 & 25.03 & 21.96 & 28.40 & 27.83 & 17.67 & 18.30 & 18.59 & 21.57 & 4.73 \\
Stereo + SparseGS & 16.05 & 24.51 & 20.86 & 27.93 & 27.22 & 17.35 & 18.68 & 16.59 & 21.15 & 4.49 \\
\bottomrule
\end{tabular}}
\vspace{0.15em}

\textbf{mSSIM$\uparrow$}\\[1pt]
\resizebox{\linewidth}{!}{%
\begin{tabular}{lcccccccccc}
\toprule
Method & Action Figure & Ball & Chicken & Dog & Espresso & Optics & Salt Pepper & Shelf & Avg & Std \\
\midrule
\multicolumn{11}{l}{\textit{Light field capture}} \\
Monocular + 3DGS & 0.530 & 0.689 & 0.677 & 0.656 & 0.695 & 0.577 & 0.720 & 0.703 & 0.656 & 0.063 \\
Monocular + FSGS & 0.526 & 0.753 & 0.759 & 0.725 & 0.710 & 0.627 & 0.766 & 0.754 & 0.703 & 0.079 \\
Monocular + SparseGS & 0.490 & 0.764 & 0.788 & 0.747 & 0.726 & 0.624 & 0.793 & 0.768 & 0.712 & 0.098 \\
Light field + 3DGS & 0.825 & 0.820 & 0.842 & 0.813 & 0.807 & 0.758 & 0.853 & 0.825 & 0.818 & 0.027 \\
Light field + FSGS & 0.717 & 0.812 & 0.843 & 0.795 & 0.805 & 0.696 & 0.852 & 0.823 & 0.793 & 0.053 \\
Light field + SparseGS & 0.813 & 0.819 & 0.846 & 0.814 & 0.806 & 0.744 & 0.854 & 0.823 & 0.815 & 0.031 \\
\midrule
\multicolumn{11}{l}{\textit{iPhone capture}} \\
Monocular + 3DGS & 0.719 & 0.779 & 0.790 & 0.795 & 0.760 & 0.690 & 0.812 & 0.728 & 0.759 & 0.040 \\
Monocular + FSGS & 0.774 & 0.854 & 0.832 & 0.865 & 0.815 & 0.724 & 0.857 & 0.868 & 0.824 & 0.048 \\
Monocular + SparseGS & 0.792 & 0.849 & 0.825 & 0.875 & 0.844 & 0.730 & 0.861 & 0.945 & 0.840 & 0.059 \\
iPhone + 3DGS & 0.689 & 0.847 & 0.818 & 0.864 & 0.781 & 0.748 & 0.837 & 0.904 & 0.811 & 0.064 \\
iPhone + FSGS & 0.801 & 0.885 & 0.841 & 0.895 & 0.855 & 0.795 & 0.863 & 0.936 & 0.859 & 0.044 \\
iPhone + SparseGS & 0.810 & 0.887 & 0.842 & 0.887 & 0.852 & 0.749 & 0.870 & 0.943 & 0.855 & 0.054 \\
\midrule
\multicolumn{11}{l}{\textit{Stereo capture}} \\
Monocular + 3DGS & 0.900 & 0.872 & 0.833 & 0.924 & 0.839 & 0.731 & 0.891 & 0.904 & 0.862 & 0.057 \\
Monocular + FSGS & 0.936 & 0.870 & 0.850 & 0.922 & 0.876 & 0.787 & 0.912 & 0.929 & 0.885 & 0.047 \\
Monocular + SparseGS & 0.916 & 0.876 & 0.839 & 0.933 & 0.835 & 0.745 & 0.891 & 0.913 & 0.868 & 0.057 \\
Stereo + 3DGS & 0.948 & 0.905 & 0.857 & 0.940 & 0.887 & 0.843 & 0.926 & 0.904 & 0.901 & 0.035 \\
Stereo + FSGS & 0.965 & 0.922 & 0.882 & 0.946 & 0.933 & 0.870 & 0.939 & 0.929 & 0.923 & 0.030 \\
Stereo + SparseGS & 0.960 & 0.915 & 0.866 & 0.943 & 0.901 & 0.852 & 0.931 & 0.912 & 0.910 & 0.034 \\
\bottomrule
\end{tabular}}
\vspace{0.15em}

\textbf{mLPIPS$\downarrow$}\\[1pt]
\resizebox{\linewidth}{!}{%
\begin{tabular}{lcccccccccc}
\toprule
Method & Action Figure & Ball & Chicken & Dog & Espresso & Optics & Salt Pepper & Shelf & Avg & Std \\
\midrule
\multicolumn{11}{l}{\textit{Light field capture}} \\
Monocular + 3DGS & 0.218 & 0.279 & 0.185 & 0.237 & 0.262 & 0.302 & 0.233 & 0.227 & 0.243 & 0.034 \\
Monocular + FSGS & 0.257 & 0.243 & 0.167 & 0.241 & 0.279 & 0.347 & 0.241 & 0.221 & 0.250 & 0.048 \\
Monocular + SparseGS & 0.221 & 0.238 & 0.162 & 0.227 & 0.256 & 0.292 & 0.224 & 0.225 & 0.231 & 0.034 \\
Light field + 3DGS & 0.183 & 0.267 & 0.166 & 0.231 & 0.267 & 0.284 & 0.245 & 0.220 & 0.233 & 0.039 \\
Light field + FSGS & 0.301 & 0.313 & 0.185 & 0.269 & 0.301 & 0.395 & 0.283 & 0.248 & 0.287 & 0.056 \\
Light field + SparseGS & 0.199 & 0.271 & 0.165 & 0.239 & 0.277 & 0.308 & 0.250 & 0.240 & 0.244 & 0.042 \\
\midrule
\multicolumn{11}{l}{\textit{iPhone capture}} \\
Monocular + 3DGS & 0.402 & 0.568 & 0.530 & 0.485 & 0.415 & 0.419 & 0.570 & 0.432 & 0.478 & 0.066 \\
Monocular + FSGS & 0.333 & 0.446 & 0.472 & 0.355 & 0.376 & 0.372 & 0.461 & 0.290 & 0.388 & 0.061 \\
Monocular + SparseGS & 0.265 & 0.447 & 0.468 & 0.310 & 0.343 & 0.327 & 0.456 & 0.215 & 0.354 & 0.088 \\
iPhone + 3DGS & 0.386 & 0.483 & 0.444 & 0.447 & 0.441 & 0.420 & 0.525 & 0.310 & 0.432 & 0.060 \\
iPhone + FSGS & 0.362 & 0.401 & 0.430 & 0.329 & 0.379 & 0.340 & 0.424 & 0.247 & 0.364 & 0.056 \\
iPhone + SparseGS & 0.260 & 0.400 & 0.423 & 0.313 & 0.339 & 0.320 & 0.433 & 0.218 & 0.338 & 0.072 \\
\midrule
\multicolumn{11}{l}{\textit{Stereo capture}} \\
Monocular + 3DGS & 0.593 & 0.377 & 0.441 & 0.346 & 0.462 & 0.533 & 0.532 & 0.430 & 0.464 & 0.078 \\
Monocular + FSGS & 0.571 & 0.377 & 0.437 & 0.353 & 0.429 & 0.510 & 0.514 & 0.421 & 0.451 & 0.069 \\
Monocular + SparseGS & 0.565 & 0.366 & 0.431 & 0.324 & 0.461 & 0.518 & 0.527 & 0.419 & 0.451 & 0.077 \\
Stereo + 3DGS & 0.478 & 0.333 & 0.420 & 0.340 & 0.434 & 0.430 & 0.458 & 0.459 & 0.419 & 0.051 \\
Stereo + FSGS & 0.432 & 0.327 & 0.406 & 0.344 & 0.419 & 0.405 & 0.429 & 0.434 & 0.399 & 0.039 \\
Stereo + SparseGS & 0.437 & 0.320 & 0.412 & 0.333 & 0.421 & 0.416 & 0.439 & 0.450 & 0.403 & 0.046 \\
\bottomrule
\end{tabular}}

\caption{Per-scene masked metrics for single-exposure captures of real-world static scenes.}
\label{tab:static-real-per-scene}
\end{table}

%% file: tables/casual-real-per-scene.tex
\begin{table}[htbp]
\centering
\scriptsize
\renewcommand{\arraystretch}{0.95}

\textbf{mPSNR$\uparrow$}\\[2pt]
\resizebox{\linewidth}{!}{%
\begin{tabular}{lcccccccccc}
\toprule
Method & Action Figure & Ball & Chicken & Dog & Espresso & Optics & Salt Pepper & Shelf & Avg & Std \\
\midrule
\multicolumn{11}{l}{\textit{iPhone Capture}} \\
Monocular & 21.10 & 19.67 & 23.33 & 20.55 & 25.26 & 21.70 & 20.98 & 21.41 & 21.75 & 1.65 \\
Monocular w/ Prior & 25.29 & 21.24 & 25.54 & 21.63 & 24.75 & 23.04 & 23.13 & 24.76 & 23.67 & 1.55 \\
iPhone & 19.89 & 21.82 & 22.30 & 22.86 & 22.44 & 20.19 & 22.15 & 25.31 & 22.12 & 1.57 \\
iPhone w/ Prior & 24.21 & 24.05 & 23.06 & 20.78 & 24.09 & 19.95 & 23.19 & 26.69 & 23.25 & 1.97 \\
\addlinespace[2pt]
\midrule
\addlinespace[2pt]
\multicolumn{11}{l}{\textit{Stereo Capture}} \\
Monocular & 29.33 & 32.19 & 27.83 & 23.68 & 36.60 & 25.37 & 24.43 & 23.71 & 27.89 & 4.33 \\
Monocular w/ Prior & 31.15 & 32.71 & 26.37 & 29.10 & 35.59 & 25.66 & 22.98 & 27.86 & 28.93 & 3.83 \\
Stereo & 30.24 & 32.19 & 26.55 & 26.67 & 36.29 & 26.05 & 23.19 & 26.33 & 28.44 & 3.93 \\
Stereo w/ Prior & 30.80 & 32.77 & 26.70 & 28.81 & 35.78 & 25.70 & 23.62 & 27.63 & 28.97 & 3.72 \\
\bottomrule
\end{tabular}}
\vspace{0.3em}

\textbf{mSSIM$\uparrow$}\\[2pt]
\resizebox{\linewidth}{!}{%
\begin{tabular}{lcccccccccc}
\toprule
Method & Action Figure & Ball & Chicken & Dog & Espresso & Optics & Salt Pepper & Shelf & Avg & Std \\
\midrule
\multicolumn{11}{l}{\textit{iPhone Capture}} \\
Monocular & 0.707 & 0.800 & 0.848 & 0.846 & 0.877 & 0.765 & 0.844 & 0.886 & 0.822 & 0.057 \\
Monocular w/ Prior & 0.810 & 0.849 & 0.895 & 0.875 & 0.879 & 0.828 & 0.874 & 0.920 & 0.866 & 0.033 \\
iPhone & 0.613 & 0.824 & 0.843 & 0.848 & 0.841 & 0.752 & 0.850 & 0.909 & 0.810 & 0.085 \\
iPhone w/ Prior & 0.765 & 0.864 & 0.860 & 0.793 & 0.871 & 0.795 & 0.859 & 0.924 & 0.841 & 0.049 \\
\addlinespace[2pt]
\midrule
\addlinespace[2pt]
\multicolumn{11}{l}{\textit{Stereo Capture}} \\
Monocular & 0.919 & 0.943 & 0.912 & 0.892 & 0.965 & 0.878 & 0.869 & 0.907 & 0.911 & 0.030 \\
Monocular w/ Prior & 0.939 & 0.952 & 0.938 & 0.947 & 0.967 & 0.911 & 0.893 & 0.951 & 0.937 & 0.022 \\
Stereo & 0.933 & 0.945 & 0.917 & 0.927 & 0.965 & 0.890 & 0.866 & 0.934 & 0.922 & 0.029 \\
Stereo w/ Prior & 0.936 & 0.953 & 0.938 & 0.947 & 0.968 & 0.913 & 0.897 & 0.950 & 0.937 & 0.021 \\
\bottomrule
\end{tabular}}
\vspace{0.3em}

\textbf{mLPIPS$\downarrow$}\\[2pt]
\resizebox{\linewidth}{!}{%
\begin{tabular}{lcccccccccc}
\toprule
Method & Action Figure & Ball & Chicken & Dog & Espresso & Optics & Salt Pepper & Shelf & Avg & Std \\
\midrule
\multicolumn{11}{l}{\textit{iPhone Capture}} \\
Monocular & 0.272 & 0.272 & 0.175 & 0.204 & 0.224 & 0.278 & 0.251 & 0.196 & 0.234 & 0.037 \\
Monocular w/ Prior & 0.137 & 0.209 & 0.127 & 0.152 & 0.223 & 0.193 & 0.197 & 0.150 & 0.174 & 0.034 \\
iPhone & 0.320 & 0.269 & 0.207 & 0.206 & 0.276 & 0.306 & 0.258 & 0.170 & 0.251 & 0.049 \\
iPhone w/ Prior & 0.187 & 0.196 & 0.171 & 0.197 & 0.223 & 0.241 & 0.219 & 0.146 & 0.198 & 0.029 \\
\addlinespace[2pt]
\midrule
\addlinespace[2pt]
\multicolumn{11}{l}{\textit{Stereo Capture}} \\
Monocular & 0.129 & 0.128 & 0.181 & 0.268 & 0.170 & 0.205 & 0.312 & 0.246 & 0.205 & 0.062 \\
Monocular w/ Prior & 0.123 & 0.126 & 0.191 & 0.151 & 0.216 & 0.175 & 0.335 & 0.192 & 0.189 & 0.063 \\
Stereo & 0.118 & 0.127 & 0.191 & 0.183 & 0.186 & 0.186 & 0.325 & 0.213 & 0.191 & 0.059 \\
Stereo w/ Prior & 0.127 & 0.126 & 0.188 & 0.149 & 0.215 & 0.171 & 0.335 & 0.198 & 0.189 & 0.063 \\
\bottomrule
\end{tabular}}

\caption{Per-scene masked metrics for reconstruction from casual video input, including reconstructing with learned priors.}
\label{tab:casual-real-per-scene}
\end{table}

%% file: tables/dynamic-real-per-scene.tex
\begin{table}[htbp]
\centering
\scriptsize
\renewcommand{\arraystretch}{0.95}

\resizebox{\linewidth}{!}{%
\begin{tabular}{lcccccccccc}
\multicolumn{11}{c}{\textbf{PSNR$\uparrow$}} \\
\toprule
Method & Ball & Coffee & Orange & Roll & Spinner & Sugar & Scissors & Tissue & Avg & Std \\
\midrule
\multicolumn{11}{l}{\textit{iPhone Capture}} \\
Monocular & 22.43 & 19.85 & 26.60 & 31.44 & 28.09 & 17.88 & 23.56 & 26.24 & 24.51 & 4.17 \\
iPhone & 23.93 & 31.55 & 27.59 & 40.15 & 35.08 & 25.51 & 35.45 & 34.40 & 31.71 & 5.24 \\
\midrule
\addlinespace[2pt]
\multicolumn{11}{l}{\textit{Stereo Capture}} \\
Monocular & 26.18 & 36.89 & 27.71 & 36.43 & 19.78 & 31.57 & 15.99 & 38.82 & 29.17 & 7.78 \\
Stereo & 24.42 & 34.86 & 31.24 & 38.12 & 38.09 & 34.80 & 38.26 & 36.02 & 34.48 & 4.40 \\
\bottomrule
\end{tabular}}
\vspace{0.3em}

\resizebox{\linewidth}{!}{%
\begin{tabular}{lcccccccccc}
\multicolumn{11}{c}{\textbf{SSIM$\uparrow$}} \\
\toprule
Method & Ball & Coffee & Orange & Roll & Spinner & Sugar & Scissors & Tissue & Avg & Std \\
\midrule
\multicolumn{11}{l}{\textit{iPhone Capture}} \\
Monocular & 0.855 & 0.844 & 0.935 & 0.961 & 0.930 & 0.628 & 0.929 & 0.945 & 0.879 & 0.103 \\
iPhone & 0.894 & 0.938 & 0.906 & 0.976 & 0.961 & 0.750 & 0.969 & 0.966 & 0.920 & 0.070 \\
\midrule
\addlinespace[2pt]
\multicolumn{11}{l}{\textit{Stereo Capture}} \\
Monocular & 0.930 & 0.960 & 0.941 & 0.968 & 0.939 & 0.927 & 0.874 & 0.974 & 0.939 & 0.030 \\
Stereo & 0.917 & 0.956 & 0.949 & 0.974 & 0.973 & 0.931 & 0.968 & 0.970 & 0.955 & 0.020 \\
\bottomrule
\end{tabular}}
\vspace{0.3em}

\resizebox{\linewidth}{!}{%
\begin{tabular}{lcccccccccc}
\multicolumn{11}{c}{\textbf{LPIPS$\downarrow$}} \\
\toprule
Method & Ball & Coffee & Orange & Roll & Spinner & Sugar & Scissors & Tissue & Avg & Std \\
\midrule
\multicolumn{11}{l}{\textit{iPhone Capture}} \\
Monocular & 0.309 & 0.345 & 0.249 & 0.147 & 0.220 & 0.532 & 0.237 & 0.201 & 0.280 & 0.111 \\
iPhone & 0.279 & 0.235 & 0.293 & 0.141 & 0.147 & 0.419 & 0.144 & 0.166 & 0.228 & 0.093 \\
\midrule
\addlinespace[2pt]
\multicolumn{11}{l}{\textit{Stereo Capture}} \\
Monocular & 0.359 & 0.288 & 0.323 & 0.236 & 0.293 & 0.434 & 0.386 & 0.241 & 0.320 & 0.065 \\
Stereo & 0.383 & 0.301 & 0.304 & 0.235 & 0.233 & 0.432 & 0.265 & 0.263 & 0.302 & 0.067 \\
\bottomrule
\end{tabular}}
\caption{Per-scene metrics for 4DGS reconstruction on real dynamic scenes.}
\label{tab:dynamic-real-per-scene}
\end{table}